\documentclass[%
superscriptaddress, 
groupedaddress,
twocolumn,
showpacs,preprintnumbers,
showkeys,
nobibnotes,
 amsmath,amssymb,
prb,
]{revtex4}

\usepackage{graphicx}
\usepackage{dcolumn}
\usepackage{bm}
\usepackage{hyperref}
\usepackage[mathlines]{lineno}
\usepackage{afterpage}
\usepackage{float}

\begin{document}

\preprint{APS/123-QED}

\title{Spin-transfer dynamics in MgO-based magnetic tunnel junctions with an out-of-plane magnetized free layer and an in-plane polarizer}

\author{E.~Kowalska} \email{e.kowalska@hzdr.de} \affiliation{Helmholtz-Zentrum Dresden - Rossendorf, Institute of Ion Beam Physics and Materials Research, Bautzner Landstrasse 400, 01328 Dresden, Germany} \affiliation{Institute of Solid State Physics, TU Dresden, Zellescher Weg 16, 01069 Dresden, Germany}
\author{V.~Sluka} \affiliation{Helmholtz-Zentrum Dresden - Rossendorf, Institute of Ion Beam Physics and Materials Research, Bautzner Landstrasse 400, 01328 Dresden, Germany} 
\author{A.~K\'{a}kay} \email{a.kakay@hzdr.de} \affiliation{Helmholtz-Zentrum Dresden - Rossendorf, Institute of Ion Beam Physics and Materials Research, Bautzner Landstrasse 400, 01328 Dresden, Germany}
\author{C.~Fowley} \affiliation{Helmholtz-Zentrum Dresden - Rossendorf, Institute of Ion Beam Physics and Materials Research, Bautzner Landstrasse 400, 01328 Dresden, Germany}
\author{J.~Lindner} \affiliation{Helmholtz-Zentrum Dresden - Rossendorf, Institute of Ion Beam Physics and Materials Research, Bautzner Landstrasse 400, 01328 Dresden, Germany}
\author{J.~Fassbender} \affiliation{Helmholtz-Zentrum Dresden - Rossendorf, Institute of Ion Beam Physics and Materials Research, Bautzner Landstrasse 400, 01328 Dresden, Germany} \affiliation{Institute of Solid State Physics, TU Dresden, Zellescher Weg 16, 01069 Dresden, Germany}
\author{A.M.~Deac} \affiliation{Helmholtz-Zentrum Dresden - Rossendorf, Institute of Ion Beam Physics and Materials Research, Bautzner Landstrasse 400, 01328 Dresden, Germany}

\date{\today}

\begin{abstract}
Here, we present an analytical and numerical model describing the magnetization dynamics in MgO-based spin-torque nano-oscillators with an in-plane magnetized polarizer and an out-of-plane free layer. We introduce the spin-transfer torque asymmetry by considering the cosine angular dependence of the resistance between the two magnetic layers in the stack. For the analytical solution, dynamics are determined by assuming a circular precession trajectory around the direction perpendicular to the plane, as set by the effective field, and calculating the energy integral over a single precession period.
In a more realistic approach, we include the bias dependence of the tunnel magnetoresistance, which is assumed empirically to be a piecewise linear function of the applied voltage. The dynamical states are found by solving the stability condition for the Jacobian matrix for out-of-plane static states. We find that the bias dependence of the tunnel magnetoresistance, which is an inseparable effect in every tunnel junction, exhibits drastic impact on the spin-torque nano-oscillator phase diagram, mainly by increasing the critical current for dynamics and quenching the oscillations at high currents. The results are in good agreement with our experimental data published elsewhere.
\end{abstract}

http://www.aip.org/publishing/pacs/pacs-alphabetical-index



\maketitle

\section{Introduction}

An electrical current passing through a ferromagnetic material gains spin polarization defined, in the first approximation, by the orientation of the magnetic moment of the layer. This collective spin-angular momentum carried by the spin-polarized current can be transferred to the~magnetic moment of a second ferromegnatic layer, thereby generating a torque on its magnetization known as spin-transfer torque (STT)\cite{Slonczewski_1996,Berger_1996}. As predicted by Slonczewski\cite{Slonczewski_1996} and Berger\cite{Berger_1996} in 1996, spin-transfer torques may be strong enough to switch the~magnetization direction of the ferromagnetic layers, without the need of external magnetic fields, as experimentally confirmed later on\cite{Tsoi_1998,Myers_1999,Katine_2000}. Furthermore, Slonczewski\cite{Slonczewski_1996} suggested that passing a~direct current (DC) through a magnetic multilayer can also induce steady-state precession of the magnetization in at least one of the ferromagnetic layers. This phenomenon was experimentally observed for the first time in metallic nano-pillars by Kiselev {\em et al.}\cite{Kiselev_2003} in 2003, in magnetic tunnel junctions (MTJs) with Al$_2$O$_3$ tunnel barriers by Petit {\em et al.}\cite{Petit_2007} in 2007, and in MgO-based MTJs by Deac {\em et al.}\cite{Deac_2008} in 2008. Spin-torque nano-oscillators (STNOs) are currently under intense investigations for their potential applications as low input power radio-frequency devices for wireless telecommunication devices, such as transmitters, receivers, mixers, phase shifters, etc.\cite{Villard_2010,Choi_2014} In comparison to conventional transistor-based electronic oscillators, STNOs offer tunability over a wide range of frequencies by adjusting the applied current\cite{Kiselev_2003,Rippard_2004}, and their lateral size can be up to 50 times smaller\cite{Rippard_2010,Villard_2010,Zeng_2013,Locatelli_2014,Choi_2014}. Simultaneously, their output powers and frequencies remain compatible with the requirements for applications, i.e. output powers in the $\mu$W range\cite{Deac_2008,Maehara_2013,Kubota_2013} and frequencies of the order of GHz\cite{Kiselev_2003,Rippard_2004,Villard_2010,Zeng_2013,Locatelli_2014,Choi_2014}.

To date, most studies focusing on spin-transfer driven dynamics were carried out on devices with both the free and the reference layers magnetized in-plane. In this configuration, under application-desirable conditions (i.e., close to zero applied fields), steady-state precession mainly occurs on clam-shell trajectories centered on the direction defined by the in-plane shape anisotropy. Consequently, only a fraction of the full magnetoresistance amplitude translates into the radio-frequency output power. 
However, when using structures with an in-plane (IP) magnetized fixed layer and an out-of-plane (OOP) magnetized free layer (a so-called hybrid geometry shown in Fig.~\ref{fig_1}(a)), the full parallel(P)-to-antiparallel (AP) resistance variation can be converted into maximized output power in the limit of 90$^\circ$ precession angle~\cite{Rippard_2010}.
Moreover, this configuration also helps to reduce the critical currents\cite{Mangin_2006} and can provide functionality regardless of the magnetic or current history~\cite{Zeng_2013,Rippard_2010,Kubota_2013,Taniguchi_2013}.
In comparison to fully metallic GMR-type devices, MgO-based tunnel junctions remain better candidates for applications as STNOs, mostly due to much higher magnetoresistance ratios, directly translating into larger output powers, and lower operation currents in the order of 1\,mA\cite{Deac_2008,Skowronski_2012,Zeng_2013} (reduced by one order of magnitude compared to the fully metallic spin valves\cite{Rippard_2010,Kiselev_2003}) leading to a significant reduction in the power consumption of the actual device.

Previous theoretical studies demonstrated that stable precession in hybrid geometry STNOs can only be sustained if the in-plane component of the spin-transfer torque (STT$_ {\|}$) exhibits an asymmetric dependence on the angle between the free and the polarizing layer. This is true for fully metallic devices, where for constant applied currents the~torque exhibits strong angular asymmetry~\cite{Slonczewski_2002}, but not for the MgO-based magnetic tunnel junctions, which do not exhibit an intrinsic asymmetry of the STT$_ {\|}$ component~\cite{Slonczewski_2005}. 
Unfortunately, the output power of the metallic STNOs (in the order of 0.1\,nW) is not sufficient for most applications. However, recent experimental reports showed that spin-transfer driven dynamics can also be sustained in similarly designed MgO-based MTJs\cite{Kubota_2013,Deac_2008,Zeng_2013,Maehara_2013,Skowronski_2012}, exhibiting output powers up to 0.55\,$\mu$W\cite{Kubota_2013}, in spite of the lack of STT$_{\|}$ angular asymmetry\cite{Slonczewski_2002}.

These results have so far been interpreted by defining the angular asymmetry of STT$_ {\|}$ based on the angle-dependent tunneling resistivity function suggested by Slonczewski\cite{Slonczewski_2007,Slonczewski_2005}. This formalism is analogues to the one used so far only for metallic GMR structures~\cite{Taniguchi_2013}, i.e. defined by some asymmetry constants.
In this paper, we suggest phenomenological and straightforward explanation of the mechanism for sustaining steady-state precession in hybrid geometry MgO-based MTJs, defining an angular asymmetry of STT$_ {\|}$ with measurable parameters, which makes this model potentially more suitable for comparisons with experimental data.
This mechanism is based on the strong cosine-type angular dependence of the tunnel magnetoresistance which, at constant applied current, translates into an angle-dependent voltage component, giving rise to the angular asymmetry of STT$_ {\|}$ and, thus, enabling steady-state precession to be sustained. We analytically solve the Landau-Lifshitz-Gilbert-Slonczewski (LLGS) equation for a typical device with circular cross-section, under perpendicular applied fields and currents. We assume that the magnetization precesses along a circular trajectory around the direction perpendicular to the plane, as set by the effective field\cite{Rippard_2010} (i.e., the crystalline anisotropy and the in-plane shape anisotropy are neglected). We also take into account the bias dependence of TMR, which has been so far neglected in similar calculations and we find that for constant currents, the bias dependence of the resistance gradually suppresses the STT$_{\|}$ angular dependence asymmetry, but it may be still sufficient to sustain precession and high output powers for relatively low values of applied currents and fields.

The here-presented results of our analytical and numerical studies compare well to our experimental results published in Ref.[23].

\section{General assumptions}

\begin{figure}[H]
\centering
\includegraphics[width=7.5cm]{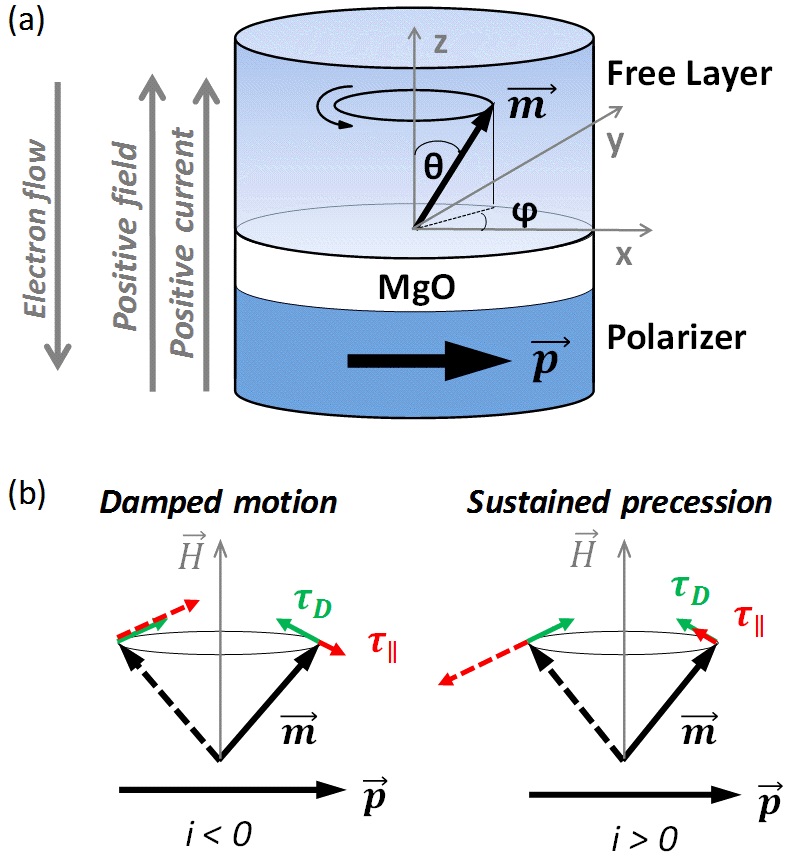}
\caption{Considered STNO geometry: (a) STNO with marked directions of positive fields and currents, (b) the principle of sustained precession (here: $\tau_{\|}$ - spin-transfer torque, $\tau_D$ - damping torque).}
\label{fig_1}
\end{figure}

In magnetoresistive multilayers, the in-plane component of STT is responsible for counteracting the damping torque and, thus, sustaining steady-state precession of the magnetization in the free layer. In magnetic tunnel junctions, the magnitude of STT$_{\|}$ depends on the angle $\beta$ between the~magnetizations of the free and reference layers\cite{Slonczewski_1989,Moodera_1996} (marked as ${\bf m}$ and ${\bf p}$ vectors in Fig.~\ref{fig_1}), as well as the magnitude of the applied voltage, $V$\cite{Slonczewski_2007,Theodonis_2006,Kubota_2008}. However, since the applied current, $I_{DC}$, is assumed to be constant, the magnitude of STT$_{\|}$ is then directly proportional to the variation of the junction resistance, $R$, according to the following formula:

\begin{equation}
\begin{split}
|STT_{\|}| = & |\frac{\partial \tau_{\|}}{\partial V} V [ {\bf m} \times ({\bf m} \times {\bf p}) ]| \propto |R I_{DC} \sin(\beta)|,
\end{split}
\label{eq_STT_in_plane}
\end{equation}

where $\partial \tau_{\|}/\partial V[T/V]$ is a torkance, i.e. derivative of an in-plane component of the spin-transfer torque with respect to the voltage\cite{Kubota_2008,Theodonis_2006}.

Investigated STNO geometry is presented in Fig.~\ref{fig_1}(a). Similarly to Rippard {\em et al.}\cite{Rippard_2010}, we assume a circular magnetization precession trajectory around the OOP axis, with a constant precession angle $\theta$ for a given applied current. As the magnetization of the free layer precesses around the $z$-axis, the angle between the magnetic moments of the two layers changes and, thus, through the magnetoresistance effect, the voltage also changes if the experiment is conducted at constant applied current. To be more specific, during the oscillation, the resistance varies between a maximum and a minimum value while approaching the P and AP configurations (i.e., for $\beta_{max}$ and $\beta_{min}$, which corresponds to $\varphi\,\approx\,$0 and $\varphi\,\approx\,\pi$, respectively; see Fig.~\ref{fig_1}(a)), following a cosine-type angular dependence of the tunnel magnetoresistance\cite{Slonczewski_1989,Moodera_1996}. This assumption, i.e. $R \propto \cos(\varphi)$, effectively introduces a spin-torque angular dependence asymmetry into the equation~\ref{eq_STT_in_plane} (as also schematically shown in Fig.~\ref{fig_1}(b)).

The motion of the free layer magnetization ${\bf m}$ is described by the LLGS equation\cite{Slonczewski_1996}:
 
\begin{equation}
\begin{split}
\frac{d{\bf m}}{dt}& = - \gamma \mu_0(H_{ext} + H_{k_{\bot}} m_z)({\bf m} \times {\bf n_z}) \\ & + \alpha ({\bf m} \times \frac{d{\bf m}}{dt}) + \gamma \frac{\partial \tau_{\|}}{\partial V} V [ {\bf m} \times ({\bf m} \times {\bf n_x}) ] \, ,
\end{split}
\label{eq_1}
\end{equation}

where $\mu_0 H_{ext}$ is an applied magnetic field, $\alpha$ is a Gilbert damping constant, $\mu_0 H_{k_{\bot}}$ is an effective magnetization along the OOP direction ($\mu_0 H_{k_{\bot}} = \mu_0 H_k - \mu_0 M_s$, where $\mu_0 H_k$ is a magnetic anisotropy field and $M_s$ is saturation magnetization), ${\bf n_{x}}$ and ${\bf n_{z}}$ are the unit vectors of the coordinate system presented in Fig.~\ref{fig_1}(a). It is worth to note that ${\bf n_{x}}$ is, in fact, a unit vector along the direction of the fixed layer magnetization, marked as the {\bf p} vector in Fig.~\ref{fig_1}(a).

The perpendicular (field-like) component of spin-transfer torque is excluded from eq.~\ref{eq_1}, since its influence on static and dynamic states in the system was found to be negligible compared to the effective field acting along $z$-axis\cite{Kubota_2008} for realistic values of the $\partial^2 \tau_{\bot} / \partial V^2$ constant (i.e., a torkance of the perpendicular component of STT). For instance, as reported by Kubota {\em et al.}~\cite{Kubota_2008} for MgO-based MTJs, $\partial^2 \tau_{\bot} / \partial V^2$\,=\,1.8\,mT/V$^2$ is at least one order of magnitude smaller compared to the~in-plane torkance $\partial \tau_{\|} / \partial V$\,=\,12.5\,mT/V for typically used voltages up to 1\,V.

To sustain the steady-state precession, the energy supplied by the in-plane spin-torque term, $STT_{\|} = \gamma [(\partial \tau_{\|}) / \partial V] V [ {\bf m} \times ({\bf m} \times {\bf p}) ]$, and energy dissipated through the damping, $\alpha ({\bf m} \times (d{\bf m} / dt))$, must compensate over a full precession period, $T$, which is described by the following integral:

\begin{equation}
\int_0^T \Big\{ \alpha ({\bf m} \times \frac{d{\bf m}}{dt}) + \gamma \frac{\partial \tau_{\|}}{\partial V} V [{\bf m} \times ({\bf m} \times {\bf p})] \Big\} dt = 0.
\label{eq_2}
\end{equation}

In the analytical calculations presented in section~\ref{Angular dependence of the TMR as a mechanism for sustained precession}), we introduce the STT asymmetry by considering the cosine angular dependence of the resistance, and derive the necessary conditions for dynamics from the energy integral shown above (eq.~\ref{eq_2}).

In the section~\ref{Influence of the TMR bias dependence}, the~bias dependence of the tunnel magnetoresistance is additionally taken into account. 
In MgO-based MTJs, the TMR exhibits a maximum at zero-bias, and then gradually decreases when increasing the magnitude of the bias voltage~\cite{Moodera_1996,Zhang_1997,Han_2001,Slonczewski_2005,Gao_2007,Kalitsov_2013}. Since the resistance variation with the bias for the P state is usually significantly smaller compared to the AP state, we assume the P state resistance to remain constant within the usable voltage range and the AP state resistance as a linear function of the~applied voltage. Note that such TMR bias dependence is just a linear approximation of the real TMR versus voltage function, which is, in fact, linear only at a temperature of 0\,K, and deviates from the linear dependence at the range of low voltage with increasing ambient temperature\cite{Moodera_1996,Zhang_1997}. This may  not influence STNO dynamics, since this low voltage range is usually below the onset value for precession.
It is also worth to note that, typically, the experimental bias voltage dependence of the TMR is asymmetric with respect to zero-bias~\cite{Yuasa_2004,Slonczewski_2005,Kalitsov_2013,Moodera_1996,Gao_2007}. It is, however, not essential in here-presented calculations, since dynamics in hybrid geometry MTJs occur only  for one particular current/voltage sign.

\section{Dynamic and static phase diagrams}

\subsection{Steady-state precession allowed by angular dependence of TMR}
\label{Angular dependence of the TMR as a mechanism for sustained precession}

\subsubsection{Steady-state precession}

Since in magnetic tunnel junctions the STT$_{\|}$ depends on the voltage across the~barrier, we convert the applied current $I_{DC}$ into the voltage $V$ with the following formula:

\begin{equation}
V = R I_{DC} = \big[ R_P + \frac{1}{2} \Delta R_0 (1 - {\bf m} \cdot {\bf p}) \big] I_{DC}.
\label{eq_3_1}
\end{equation}

Here, $R$ is the resistance obtained for a given voltage value at a given point on the precession trajectory, $R_P$ is the resistance of the parallel state, $\Delta R_0$ is the resistance difference between the P and AP states close to zero bias, and $({\bf m} \cdot {\bf p})$ is the projection of the free layer magnetization vector on the polarization direction, i.e. the magnetization component contributing to TMR. 

Taking into account the cosine angular dependence of TMR, and neglecting the bias dependence of TMR, leads to the following system of equations for the three degrees of freedom of the Cartesian coordinate system:

\begin{equation}
\begin{cases}
& m_x \rightarrow \int_0^T \Big\{ \alpha (m_y \dot{m_z} - m_z \dot{m_y}) \\ & - \gamma \frac{\partial \tau_{\|}}{\partial V} I_{DC} \big[ R_P + \frac{1}{2} \Delta R_0 (1 - m_x) \big] \big( {m_y}^2 + {m_z}^2 \big) \Big\} dt \\ & = 0 \\
\\
& m_y \rightarrow \int_0^T \Big\{ \alpha (m_z \dot{m_x} - m_x \dot{m_z}) \\ & + \gamma \frac{\partial \tau_{\|}}{\partial V} I_{DC} \big[ R_P + \frac{1}{2} \Delta R_0 (1 - m_x) \big] m_x m_y \Big\} dt = 0 \\
\\
& m_z \rightarrow \int_0^T \Big\{ \alpha (m_x \dot{m_y} - m_y \dot{m_x}) \\ & - \gamma \frac{\partial \tau_{\|}}{\partial V} I_{DC} \big[ R_P + \frac{1}{2} \Delta R_0 (1 - m_x) \big] m_x m_z \Big\} dt = 0 \\
\end{cases}
\label{eq_7_1}
\end{equation}

Subsequently, equations~\ref{eq_7_1} are transferred to the spherical coordinate system in order to simplify the description of an out-of-plane precessional state occurring at given values of applied fields and currents (namely, from now on, it will be described only by the~precession angle $\theta$):

\begin{equation}
\begin{cases}
&\int_0^{\frac{2\pi}{\omega}}\Big\{-\alpha \omega \sin{\theta} \cos{\theta} \cos{\omega t} \\ & - \gamma \frac{\partial \tau_{\|}}{\partial V} I_{DC} \big[ R_P + \frac{1}{2} \Delta R_0 (1 - \sin{\theta} \cos{\omega t}) \big] \\ & \big( {\sin}^2{\theta} {\sin}^2{\omega t} + {\cos}^2{\theta} \big) \Big\} dt = 0 \\
\\
&\int_0^{\frac{2\pi}{\omega}}\Big\{-\alpha \omega \sin{\theta} \cos{\theta} \sin{\omega t} \\ & + \gamma \frac{\partial \tau_{\|}}{\partial V} I_{DC} \big[ R_P + \frac{1}{2} \Delta R_0 (1 - \sin{\theta} \cos{\omega t}) \big] \\ & {\sin}^2{\theta} \sin{\omega t} \cos{\omega t} \Big\} dt = 0 \\
\\
&\int_0^{\frac{2\pi}{\omega}}\Big\{ \alpha \omega {\sin}^2{\theta} \\ & - \gamma \frac{\partial \tau_{\|}}{\partial V} I_{DC} \big[ R_P + \frac{1}{2} \Delta R_0 (1 - \sin{\theta} \cos{\omega t}) \big] \\ & \sin{\theta} \cos{\theta} \cos{\omega t} \Big\} dt = 0 \\
\end{cases}
\label{eq_8_1}
\end{equation}

The above-presented integrals are solved for a full precession period (i.e., the time range from 0 to $2\pi / \omega$). Solving the first and the second integrals leads us to contradiction equations. The lack of solutions for these two integrals makes sense also from the~physics point of view, i.e. since the integrals of the damping torque over a single precession cycle of $x$ and $y$ components are equal to zero, the equivalent integrals of the in-plane STT term have to be zero too (note that the damping and STT$_{\|}$ balance each other). A non-zero result is then expected only for the integration along the $z$-axis alone. Solving the third integral leads to the following equation:

\begin{equation}
4\alpha + \frac{\partial \tau_{\|}}{\partial V} I_{DC} \Delta R_0 \cos\theta \frac{\gamma}{\omega}=0.
\label{eq_new_1}
\end{equation}

Implementing the formula for the Larmor frequency~\cite{Coey_2010} ($\omega = -\gamma B$) enables us to incorporate the magnetic field $\mu_0 H = \mu_0 H_{ext} + \mu_0 H_{k_{\bot}}\cos\theta$ into the eq.~\ref{eq_new_1}:

\begin{equation}
4\alpha (\mu_0 H_{ext} + \mu_0 H_{k_{\bot}}\cos\theta) = \frac{\partial \tau_{\|}}{\partial V} I_{DC} \Delta R_0 \cos\theta.
\label{eq_new_2}
\end{equation}

In the final step of calculations, the onset angle for precession, $\theta_{onset}$, is estimated in order to find the boundaries of the dynamical region.
We assume that $\theta_{onset}$ is the smallest angle where precession can be sustained, i.e. imposing $\theta \rightarrow 0$ for positive applied fields and $\theta \rightarrow \pi$ for negative applied fields. This assumption leads us to the following relation between the critical current for steady-state precession and the external field:

\begin{equation}
\mu_0 H_{ext}^{\theta \rightarrow 0} (I_{DC}) = \pm \frac{ \frac{\partial \tau_{\|}}{\partial V} \Delta R_0}{4 \alpha} I_{DC} \mp \mu_0 H_{k_{\bot}}.
\label{eq_9_1}
\end{equation}

Here, $\mu_0 H_{ext(\theta_{onset})} (I_{DC}) = (\frac{\partial \tau_{\|}}{\partial V} \Delta R_0 / 4 \alpha) I_{DC} - \mu_0 H_{k_{\bot}}$ refers to $\theta \rightarrow 0$ for positive fields applied along $+z$ direction, and $\mu_0 H_{ext(\theta_{onset})} (I_{DC}) = - (\frac{\partial \tau_{\|}}{\partial V} \Delta R_0 / 4 \alpha) I_{DC} + \mu_0 H_{k_{\bot}}$ is the solution for $\theta \rightarrow \pi$ for negative fields applied along $-z$ direction.
Both functions (i.e., for $\theta \rightarrow 0$ and $\theta \rightarrow \pi$) are plotted as solid lines in Fig.~\ref{Diagram_dRdV_0}. Stable out-of-plane dynamics are expected to occur in the region above the lines. The~presence of dynamics only for positive currents was as expected, since only for this particular current direction (i.e., current favouring the AP state), the in-plane spin-transfer torque is efficient in overcoming the damping~\cite{Taniguchi_2013,Guo_2015} (see the scheme of the~"Sustained precession" case in Fig.~\ref{fig_1}(b)).

An analytical solution for the large angle regime (i.e., when the magnetization vector precesses in the plane of the free layer) could not be found using this calculation method, since at the limit of $\theta\,\rightarrow\,90^{\circ}$ the critical current and field approach infinity. This is due to the fact that for $\theta\,=\,90^{\circ}$ the $z$ component of the magnetization is equal zero and, thus, the integration along the $z$-axis looses its physical sense.

\subsubsection{Static states}

For small applied fields and relatively high currents, STT$_{\|}$ may become large enough to stabilize a static in-plane state in the free layer. Note that the magnetization of the free layer tends to stabilize in the sample plane, since the polarizer supplies an in-plane polarization of the flowing electrons.
We start from the initial condition for static states, $\frac{d{\bf m}}{dt}$\,=\,0 (i.e., when eq.~\ref{eq_1} is equal to zero), expressed for all three degrees of freedom of the Cartesian coordinate system:

\begin{equation}
\begin{cases}
&\frac{d{\bf {m_x}}}{dt} = - \gamma (\mu_0 H_{ext} + \mu_0 H_{k_{\bot}} m_z) m_y \\ & + \alpha (m_y \dot{m_z} - m_z \dot{m_y}) \\ & - \gamma \frac{\partial \tau_{\|}}{\partial V} I_{DC} \big[ R_P + \frac{1}{2} \Delta R_0 (1 - m_x) \big] \big( {m_y}^2 + {m_z}^2 \big) = 0 \\
\\
&\frac{d{\bf {m_y}}}{dt} = \gamma (\mu_0 H_{ext} + \mu_0 H_{k_{\bot}} m_z) m_x \\ & + \alpha (m_z \dot{m_x} - m_x \dot{m_z}) \\ & + \gamma \frac{\partial \tau_{\|}}{\partial V} I_{DC} \big[ R_P + \frac{1}{2} \Delta R_0 (1 - m_x) \big] m_x m_y = 0 \\
\\
&\frac{d{\bf {m_z}}}{dt} = \alpha (m_x \dot{m_y} - m_y \dot{m_x}) \\ &  + \gamma \frac{\partial \tau_{\|}}{\partial V} I_{DC} \big[ R_P + \frac{1}{2} \Delta R_0 (1 - m_x) \big] m_x m_z = 0 \\
\end{cases}
\label{eq_7_1_2}
\end{equation}

Subsequently, the equations \ref{eq_7_1_2} are expressed with spherical coordinates, as follows:

\begin{equation}
\begin{cases}
&-\gamma (\mu_0 H_{ext} + \mu_0 H_{k_{\bot}} \cos{\theta}) \sin{\theta}\sin{\omega t} \\ & - \alpha \omega \sin{\theta} \cos{\theta} \cos{\omega t} \\ & - \gamma \frac{\partial \tau_{\|}}{\partial V} I_{DC} \big[ R_P + \frac{1}{2} \Delta R_0 (1 - \sin{\theta} \cos{\omega t}) \big] \\ & \big( {\sin}^2{\theta} {\sin}^2{\omega t} + {\cos}^2{\theta} \big) = 0 \\
\\
&\gamma (\mu_0 H_{ext} + \mu_0 H_{k_{\bot}} \cos{\theta}) \sin{\theta}\cos{\omega t} \\ & - \alpha \omega \sin{\theta} \cos{\theta} \sin{\omega t} \\ & + \gamma \frac{\partial \tau_{\|}}{\partial V} I_{DC} \big[ R_P + \frac{1}{2} \Delta R_0 (1 - \sin{\theta} \cos{\omega t}) \big] \\ & {\sin}^2{\theta} \sin{\omega t} \cos{\omega t} = 0 \\
\\
&\alpha \omega {\sin}^2{\theta} + \gamma \frac{\partial \tau_{\|}}{\partial V} I_{DC} \big[ R_P + \frac{1}{2} \Delta R_0 (1 - \sin{\theta} \cos{\omega t}) \big] \\ & \sin{\theta} \cos{\theta} \cos{\omega t} = 0 \\
\end{cases}
\label{eq_8_1_2}
\end{equation}

Imposing that the magetization should turn to an in-plane state (defined by $\theta$\,=\,$\frac{\pi}{2}$, which indicates that an actual in-plane direction is, at this point, unknown), leads to a~following formula:

\begin{equation}
\mu_0 H_{ext} + \frac{\partial \tau_{\|}}{\partial V} I_{DC} \big[ R_P + \frac{1}{2} \Delta R_0 (1 - \cos{\omega t}) \big] \sin{\omega t} = 0.
\label{eq_9_1_2}
\end{equation}

Considering that $|\sin{\omega t}| \leqslant 1$, leads to:

\begin{equation}
\Big| -\frac{\mu_0 H_{ext}}{\frac{\partial \tau_{\|}}{\partial V} I_{DC}} {\big[ R_P + \frac{1}{2} \Delta R_0 (1 - \cos{\omega t}) \big]}^{-1} \Big|  \leqslant 1.
\label{eq_10_1}
\end{equation}

Subsequently, applying the condition of $|\cos{\omega t}| \leqslant 1$ leads to the final solution for a static in-plane state of the hybrid geometry spin-torque nano-oscillator:

\begin{equation}
\mu_0 H_{ext}(I_{DC}) \leqslant \big| \frac{\partial \tau_{\|}}{\partial V} R_P I_{DC} \big|.
\label{eq_6_1}
\end{equation}

\begin{figure}[H]
\raggedleft
\includegraphics[width=9.0cm]{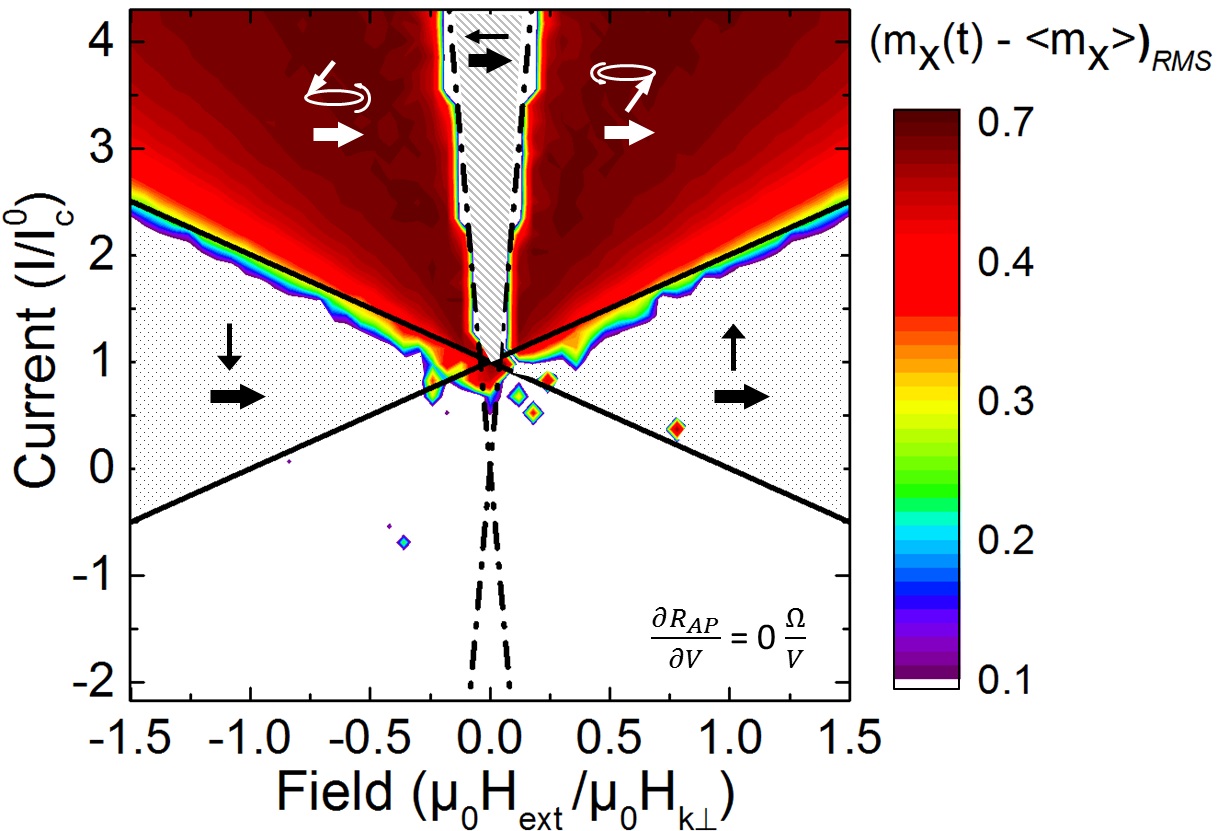}
\caption{Dynamical phase diagram of the STNO with hybrid geometry for the case when the angular dependence of the TMR is included, and the bias dependence of the TMR is neglected ($\partial R_{AP}/\partial V$\,=\,0\,$\Omega / V$). The diagram shows results of numerical integration (dynamics occurs in the coloured areas) and analytically determined onset currents (solid and dash-dot lines). The dash-dot lines (see eq.~\ref{eq_6_1}) define the region of the stable static in-plane AP state (striped area). The solid lines (see eq.~\ref{eq_9_1}) are boundaries between the region of OOP dynamics (coloured area) and the static OOP state (dotted areas). Current values, $I$, are normalized by $I_c^0$, i.e. the current value at the crossing of the critical lines for dynamics (the crossing point of the solid lines). The~magnetic field $\mu_0 H_{ext}$ is normalized by the effective out-of-plane anisotropy of the free layer ($\mu_0 H_{k_{\bot}}$). Magnetic configurations corresponding to static and dynamic states are marked with black and white arrows, respectively.}
\label{Diagram_dRdV_0}
\end{figure}

Equation \ref{eq_6_1} is plotted with dash-dot lines in Fig.~\ref{Diagram_dRdV_0}. The stability region of an in-plane state is marked as striped area. Since the in-plane state occurs only at positive applied currents (i.e., for the electron flow from the free layer to the reference layer favouring the AP state, as defined in Fig.~\ref{fig_1}(a)), the AP state is expected to be stabilized in the~striped area.
Similarly, according to symmetry argument, we preliminary assume that the region between the~dash-dot lines at the negative current range (latticed area) corresponds to the stability area of the in-plane P state.

According to the results plotted in Fig.~\ref{Diagram_dRdV_0}, the region of out-of-plane dynamics exhibits a gap at zero and small applied fields, where the out-of-plane dynamic state turns into the static AP state (see the striped area in Fig.~\ref{Diagram_dRdV_0}), as the currents required to sustain precession become infinitely large as the magnetization approaches the $xy$-plane.

\begin{figure*}[ht]
\centering
\includegraphics[width=\textwidth]{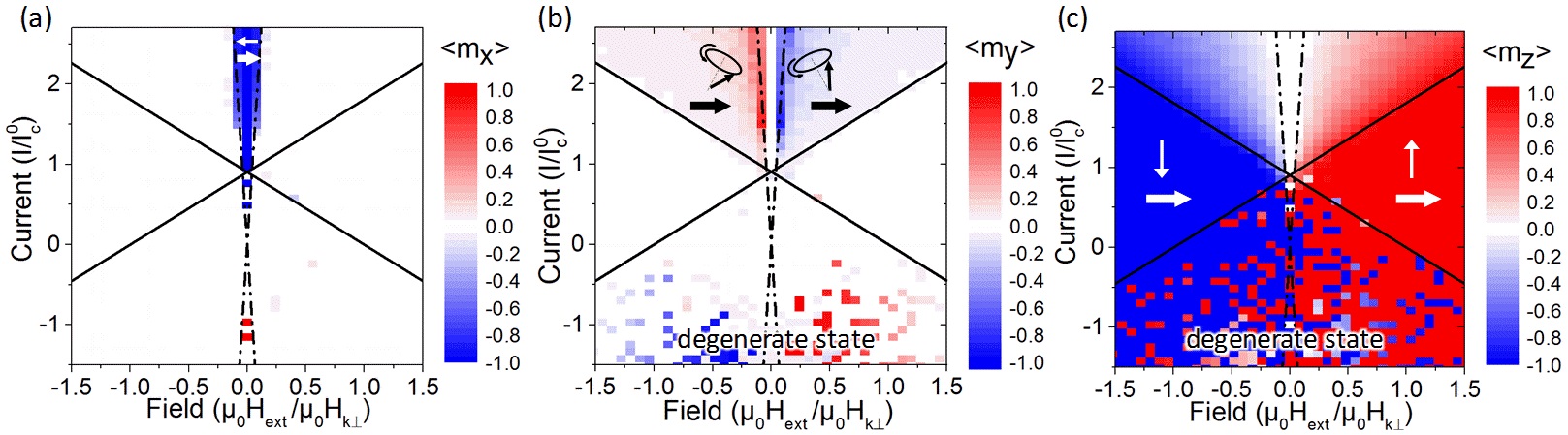}
\caption{Average magnetization components: (a) along the in-plane $x$-axis ($\langle m_x \rangle$), (b) along the in-plane $y$-axis ($\langle m_y \rangle$), and (c) along the out-of-plane $z$-axis ($\langle m_z \rangle$). The~dark blue region in (a) represents the static in-plane AP state for $\langle m_x \rangle\,=\,-1$, and is accurately defined by the analytically determined dash-dot lines (eq.~\ref{eq_6_1}). In (b), the~blue and red regions occurring at positive current range (close to the dash-dot lines) indicate a slight tilt of the magnetization precession cone towards the $+y$-axis and $-y$-axis for the negative and positive applied fields, respectively. The blue and red regions in (c) represent the static out-of-plane states for $\langle m_z \rangle\,=\,-1$ and $\langle m_z \rangle\,=\,+1$, respectively. These regions are defined by the analytically determined boundaries of the~dynamical area, i.e. the two solid lines (eq.~\ref{eq_9_1}). The static states occurring in the~area where these two stability regions overlap are degenerate. Current values, $I$, are normalized by $I_c^0$, i.e. the current value at the crossing of the critical lines for dynamics (the crossing point of the solid lines). Field values $B_{ext}$ are normalized by the effective out-of-plane anisotropy $B_{k_{\bot}}$. Magnetic configurations corresponding to static and dynamic states are marked with black and white arrows, respectively.}
\label{Diagram_dRdV_0_static}
\end{figure*}

\subsubsection{Numerical integration of LLGS equation}

The analytical solutions presented in the previous paragraphs were double-checked via numerical integration of the LLGS equation (\ref{eq_1}). To this end we used MAPLE 8 program, which enabled us to follow the evolution of the~position of the magnetization vector under a~defined set of parameters as a function of time.
The following set of parameters was used: $\partial \tau_{\|} / \partial V\,=\,0.028 T/V\,$, $\alpha\,=\,0.005$, $\Delta R_0\,=\,110\,\Omega$, $R_P\,=\,190\,\Omega$, $\mu_0 H_{k_{\bot}}\,=\,120\,mT$ (the same parameters were used for the analytical lines plotted in Fig.~\ref{Diagram_dRdV_0}). We used a simulation time of 150\,ns, while a final static or dynamic state was defined based on the last 2\,ns. The initial position of the magnetization was always set as random, so as not to overlook bi-stability regions if they occur.

The results of the numerical simulation are shown in Fig.~\ref{Diagram_dRdV_0}. Here, the colour scale represents the magnitude of the intensity of magnetization dynamics expressed by $(m_x(t)-\langle m_x \rangle)_{RMS}$, where $RMS$ is a root mean square, $m_x(t)$ is an instant $m_x$ component, and $\langle m_x \rangle$ is a~mean value of the $m_x$ component. In case of the presence of stable dynamics, the~absolute value of $(m_x(t) - \langle m_x \rangle)$ has to be greater than zero (i.e., $(m_x(t)-\langle m_x \rangle)_{RMS}$~$=$~$0$ corresponds to a lack of dynamics). Based on this definition, the maximum precession angle of $\theta$\,=\,90$^{\circ}$ (i.e., where the $m_x$ value changes sinusoidally between -1 and +1) translates into intensity of magnetization dynamics equal to 0.7 (see a dark brown region in Fig.~\ref{Diagram_dRdV_0}).
The analytically determined critical lines for dynamics (solid lines) and for the static AP state (dash-dot lines) define accurately the boundaries of the numerically obtained dynamical region (colour area).
According to the phase diagram shown in Fig.~\ref{Diagram_dRdV_0}, when the bias dependence of the magnetoresistance is ignored, similarly to the metallic case~\cite{Slonczewski_2002,Fowley_2014,Arai_2014}, stable dynamics occur only for positive current, defined as electrons flowing from the free layer to the reference layer (i.e., current favouring the AP state in Fig.~\ref{Diagram_dRdV_0}). This is a~consequence of the fact that STT$_{\|}$ is larger close to the AP state than close to the P state (see the "Sustained precession" case in Fig.~\ref{fig_1}b).

Fig.~\ref{Diagram_dRdV_0_static}(a) shows the average magnetization component along the $x$-axis as a function of applied current and field. The dark blue region indicates the presence of the~static in-plane AP state ($\langle m_x \rangle\,=\,-1$) at positive applied current in the low and zero-field range, which is in agreement with the analytically determined region of the AP state stability (striped area in Fig.~\ref{Diagram_dRdV_0}). The presence of stability area of the P state, also predicted analytically (see latticed area in Fig.~\ref{Diagram_dRdV_0}), is not confirmed with the numerically obtained data in the investigated current range. For a random initial state, the region between the dash-dot lines at the negative current range is characterized by various static states; namely, by the in-plane P state, i.e. for $\langle m_x \rangle\,=\,+1$ (shown in Fig.~\ref{Diagram_dRdV_0_static}(a)), and the two out-of-plane states, i.e. for $\langle m_z \rangle\,=\,+1$ and $\langle m_z \rangle\,=\,-1$ (see Fig.~\ref{Diagram_dRdV_0_static}(c)).

The average magnetization along the in-plane $y$-axis is plotted in Fig.~\ref{Diagram_dRdV_0_static}(b). The~blue and red regions appearing between the solid lines at positive current range (i.e., in the region of dynamics) indicate a slight tilt of the magnetization precession cone towards the $y$-axis, which gradually increases while approaching the stability area of the~AP state (i.e., close to the dash-dot lines). For positive applied fields and $\varphi\,\approx\,$90$^{\circ}$ the in-plane spin-transfer torque acts along the field torque, resulting in an increase of the precession speed (so also its frequency) and, thus, a decrease of the precession angle. For $\varphi\,\approx\,$270$^{\circ}$, however, the in-plane STT acting antiparallel to the field torque brings about a decrease of the precession speed (and the frequency), leading to an increase of the magnetization precession angle. This eventually leads to tilting of the magnetization precession cone towards the $-y$ direction. By analogy, at the negative applied fields, the precession cone tilts towards the $+y$ direction.

In the area where the analytical lines define the stability region of the AP state (at the dash-dot lines), both the $m_x$ and $m_y$ components are non-zero (see Fig.~\ref{Diagram_dRdV_0_static}(a) and (b), respectively). Moreover, according to the corresponding dynamical diagram, shown in Fig.~\ref{Diagram_dRdV_0}, there is no out-of-plane dynamics in the region exactly at the dash-dot lines. All these arguments lead us to the conclusion that, between the~stability area of the~AP state and the area of OOP steady-state precession, there is a~transition region of the static canted state, where the magnetization is tilted from the~$-x$ direction towards $y$-axis.

Fig.~\ref{Diagram_dRdV_0_static}(c) shows the average magnetization components along the $z$-axis. The~blue and red regions indicate the presence of static out-of-plane states in the system (i.e., for $\langle m_z \rangle \,=\,-1$ and $\langle m_z \rangle \,=\,+1$, respectively), which can be also defined with the~analytical solution for the onset current for dynamics (solid lines), according to eq.~\ref{eq_9_1}. This means that for currents below the onset current for dynamics, the magnetization is stabilized along the~direction of the~applied field. 
The area where these two stability regions overlap is characterized by various static states (see "degenerate state" in Fig.~\ref{Diagram_dRdV_0_static}(b) and (c)); in particular, canted states with the magnetization laying in the $yz$ plane (indicated by non-zero $\langle m_y \rangle$ and $\langle m_z \rangle$ components at negative currents in Fig.~\ref{Diagram_dRdV_0_static}(b) and (c)), or the~in-plane P state (see red points at negative currents in Fig.~\ref{Diagram_dRdV_0_static}(a)). In this region, at each point on the~diagram, a final state depends on the initial position of the magnetization (here, set as random). Thus, in order to define static states in this area, numerical simulations with defined initial states should be performed (not discussed in this paper).

\subsection{Influence of the bias dependence of TMR}
\label{Influence of the TMR bias dependence}

\subsubsection{Steady-state precession}

Assuming the following linear bias dependence of the resistance difference between the P and the AP states:

\begin{equation}
\Delta R = - \frac{\partial R_{AP}}{\partial V} \cdot |V| + \Delta R_0
\label{eq_4_1}
\end{equation}

for the instant angle between the magnetic moments of two layers, the expression for the voltage across the barrier (eq.~\ref{eq_3_1}) converts into:

\begin{equation}
V = I_{DC} \frac{R_P + \frac{1}{2} \Delta R_0 (1 - {\bf m} \cdot {\bf p})}{1 + \frac{1}{2} |I_{DC}| \frac{\partial R_{AP}}{\partial V} (1 - {\bf m} \cdot {\bf p})}.
\label{eq_5_1}
\end{equation}

Taking into account both the angular dependence and the bias dependence of the~TMR, the whole calculation procedure becomes much more complex and, according to our knowledge, not solvable with the previous approach (i.e., by solving the integral~\ref{eq_2}). Therefore, the LLGS equation is first expressed with the spherical coordinates:

\begin{equation}
\begin{cases}
&
\dot{\varphi} = \sin{\theta} \cdot \frac{\partial{\varphi}}{\partial t} = \frac{\gamma}{\mu_0 M_s} \frac{\partial W}{\partial \theta} + \alpha \frac{\partial{\theta}}{\partial t} \\ & + \frac{\partial \tau_{\|}}{\partial V} V \big( p_x \sin{\varphi} - p_y \cos{\varphi} \big) \\
\\
&
\dot{\theta} = -\sin{\theta} \cdot \frac{\partial{\theta}}{\partial t} = \frac{\gamma}{\mu_0 M_s} \frac{\partial W}{\partial \varphi} + \alpha {\sin{\theta}}^2 \frac{\partial{\varphi}}{\partial t} \\ & + \frac{\partial \tau_{\|}}{\partial V} V \big[ \sin{\theta} \cos{\theta} (p_x \cos{\theta} + p_y \sin{\varphi}) - p_z {\sin{\theta}}^2 \big] \\
\end{cases}
\label{eq_12_1}
\end{equation}

where:

\begin{equation}
\begin{cases}
&
\frac{\partial W}{\partial \theta} = - \mu_0 M_s \mu_0 H_{eff} \frac{\partial {\bf m}}{\partial \theta} \nonumber
\\
&\frac{\partial W}{\partial \varphi} = - \mu_0 M_s \mu_0 H_{eff} \frac{\partial {\bf m}}{\partial \varphi}
\end{cases}
\label{rotation_1}
\end{equation}

are the energy derivatives with respect to the all degrees of freedom of the system (here: $\mu_0 H_{eff}$ is the effective field).
In order to find the instability condition, for which the static out-of-plane state becomes unstable, the following equations must be fulfilled:

\begin{equation}
\begin{cases}
&tr({\bf J}) = \frac{\partial \dot{m_{\varphi}}}{\partial \varphi} + \frac{\partial \dot{m_{\theta}}}{\partial \theta} < 0  \\
\\
&det({\bf J}) = \frac{\partial \dot{m_{\varphi}}}{\partial \varphi} \cdot \frac{\partial \dot{m_{\theta}}}{\partial \theta} - \frac{\partial \dot{m_{\theta}}}{\partial \varphi} \cdot \frac{\partial \dot{m_{\varphi}}}{\partial \theta} > 0  \\
\end{cases}
\label{eq_13_1}
\end{equation}

Here, {\bf J} is the Jacobian matrix of the system, i.e. the matrix of the first-order derivatives of function~\ref{eq_12_1}, and $tr$ and $det$ are the trace and determinant of the Jacobian matrix, respectively.

Solving the~equation \ref{eq_12_1} expressed in the coordinate system shown in Fig.~\ref{fig_1}(a) leads us to two inequalities consisting of a contradiction equation (having no solution) and an identity equation (fulfilled for all real numbers). Therefore, in order to make eq.~\ref{eq_12_1} solvable, we rotate the coordinate system in the following way: $(x,y,z) \rightarrow (x,z,-y)$. Now, in the new coordinate system, the precession angle $\theta$ is expressed as $(\pi/2 - \theta)$, while the angle in the sample plane is still defined by $\varphi$. In this way, the poles of the previous coordinate system are moved from the positions of the calculation limits (i.e., positions of $\theta = 0$ and $\theta = \pi$ are moved away from the $z$-axis), which enables us to solve eq.~\ref{eq_13_1}. The equations~\ref{rotation_1} are now expressed as:

\begin{equation}
\begin{cases}
&
\frac{\partial W}{\partial \theta} = -\mu_0 M_s  \\ &
    \left( \begin{array}{c}
      0 \\ \nonumber
      -\mu_0(H_{ext}+H_{k_{\bot}} \sin{\theta} \sin{\varphi})  \\ \nonumber
      0
    \end{array} \right) 
    \left( \begin{array}{c}
      \cos{\theta} \cos{\varphi} \\ \nonumber
      \cos{\theta} \sin{\varphi} \\ \nonumber
      -\sin{\theta}
    \end{array} \right) 
\\ & = \mu_0 M_s(\mu_0 H_{ext}+\mu_0 H_{k_{\bot}} \sin{\theta} \sin{\varphi}) \cos{\theta} \sin{\varphi} \\ \nonumber
\\
&\frac{\partial W}{\partial \varphi} = -\mu_0 M_s \\ &
    \left( \begin{array}{c}
      0 \\ \nonumber
      -\mu_0(H_{ext}+H_{k_{\bot}} \sin{\theta} \sin{\varphi})  \\ \nonumber
      0
    \end{array} \right) 
    \left( \begin{array}{c}
      -\sin{\theta} \sin{\varphi} \\ \nonumber
      \sin{\theta} \cos{\varphi} \\ \nonumber
      0
    \end{array} \right) 
\\ & = \mu_0 M_s\mu_0(H_{ext}+H_{k_{\bot}} \sin{\theta} \sin{\varphi}) \sin{\theta} \cos{\varphi} \\
\end{cases}
\end{equation}

Subsequently, the coordinate system is rotated back to its initial position, shown in Fig.~\ref{fig_1}(a), according to the following rotation: $(x,z,-y) \rightarrow (x,y,z)$. Applying the limits of $\theta \rightarrow 0$ and $\theta \rightarrow \pi$ leads us to the~final equation for the critical lines defining the region where static out-of-plane states become unstable:

\begin{equation}
\mu_0 H_{ext}^{\theta \rightarrow 0} (I_{DC}) < \Big| \frac{\frac{\partial \tau_{\|}}{\partial V}}{\alpha} \frac{ \Delta R_0 - |I_{DC}| \frac{\partial R_{AP}}{\partial V} R_P}{\big( 2 + |I_{DC}| \frac{\partial R_{AP}}{\partial V} \big)^2} I_{DC} - \mu_0 H_{k_{\bot}}\Big|.
\label{eq_xxx}
\end{equation}

\subsubsection{Static states}

Similarly to the case where the bias dependence of the resistance is neglected, for small fields and high currents, one expects STT$_{\|}$ to stabilize the static AP state in the~free layer. In order to find a solution for the AP state for the case of $\partial R_{AP}/\partial V\,>\,0$, we used a similar calculation procedure as one presented in section~\ref{Angular dependence of the TMR as a mechanism for sustained precession} (see eq.~\ref{eq_7_1_2}-\ref{eq_6_1}). Only that now the bias dependence of the resistance is additionally taken into account by incorporating the condition~\ref{eq_5_1} to eq.~\ref{eq_7_1_2}, which brings us to the following formula:

\begin{equation}
\mu_0 H_{ext} + \frac{\partial \tau_{\|}}{\partial V} I_{DC} \frac{1 + \frac{1}{2} |I_{DC}| \frac{\partial R_{AP}}{\partial V} (1 - \cos{\omega t})}{R_P + \frac{1}{2} \Delta R_0 (1 - \cos{\omega t})} \sin{\omega t} = 0.
\label{eq_9_1_2_bias}
\end{equation}

Considering that $|\sin{\omega t}| \leqslant 1$, leads to:

\begin{equation}
\Big| -\frac{\mu_0 H_{ext}}{\frac{\partial \tau_{\|}}{\partial V} I_{DC}} \frac{1 + \frac{1}{2} |I_{DC}| \frac{\partial R_{AP}}{\partial V} (1 - \cos{\omega t})}{R_P + \frac{1}{2} \Delta R_0 (1 - \cos{\omega t})} \Big|  \leqslant 1.
\label{eq_10_1_bias}
\end{equation}

Eventually, applying the condition that $|\cos{\omega t}| \leqslant 1$ brings us to the final solution for the static AP state of the hybrid STNO device:

\begin{equation}
\mu_0 H_{ext}(I_{DC}) \leqslant \big| \frac{\partial \tau_{\|}}{\partial V} R_P I_{DC} \big|.
\label{eq_6_1_bias}
\end{equation}

Note that eq.~\ref{eq_6_1} and eq.~\ref{eq_6_1_bias} are identical; namely, the~analytical solution for the static AP state remains then unchanged compared to the case when the bias dependence of the~resistance is neglected. This indicates that the bias dependence of the TMR does not influence the stability region of the AP state occurring at zero- and low applied fields.

The analytical solutions for stable out-of-plane dynamics (eq.~\ref{eq_xxx}) and the static AP state (eq.~\ref{eq_6_1_bias}), defining the region of out-of-plane dynamics, are plotted in Fig.~\ref{Diagram_dRdV_100} with solid and dash-dot lines, respectively.
Similarly to the results presented in section~\ref{Angular dependence of the TMR as a mechanism for sustained precession}, the critical lines making the onset of dynamics (solid lines) are simultaneously the boundaries of the stability regions of static out-of-plane states (dotted areas). It is, however, worth to remember that the analytical results represent a simplified case, neglecting the narrow transition area of the static canted state between the regions of the AP state and OOP dynamics (i.e., exactly at the dash-dot lines), which results from the numerical data (see description to Fig.~\ref{Diagram_dRdV_0_static}).

\subsubsection{Numerical integration of LLGS equation}

\begin{figure}[H]
\raggedleft
\includegraphics[width=8.8cm]{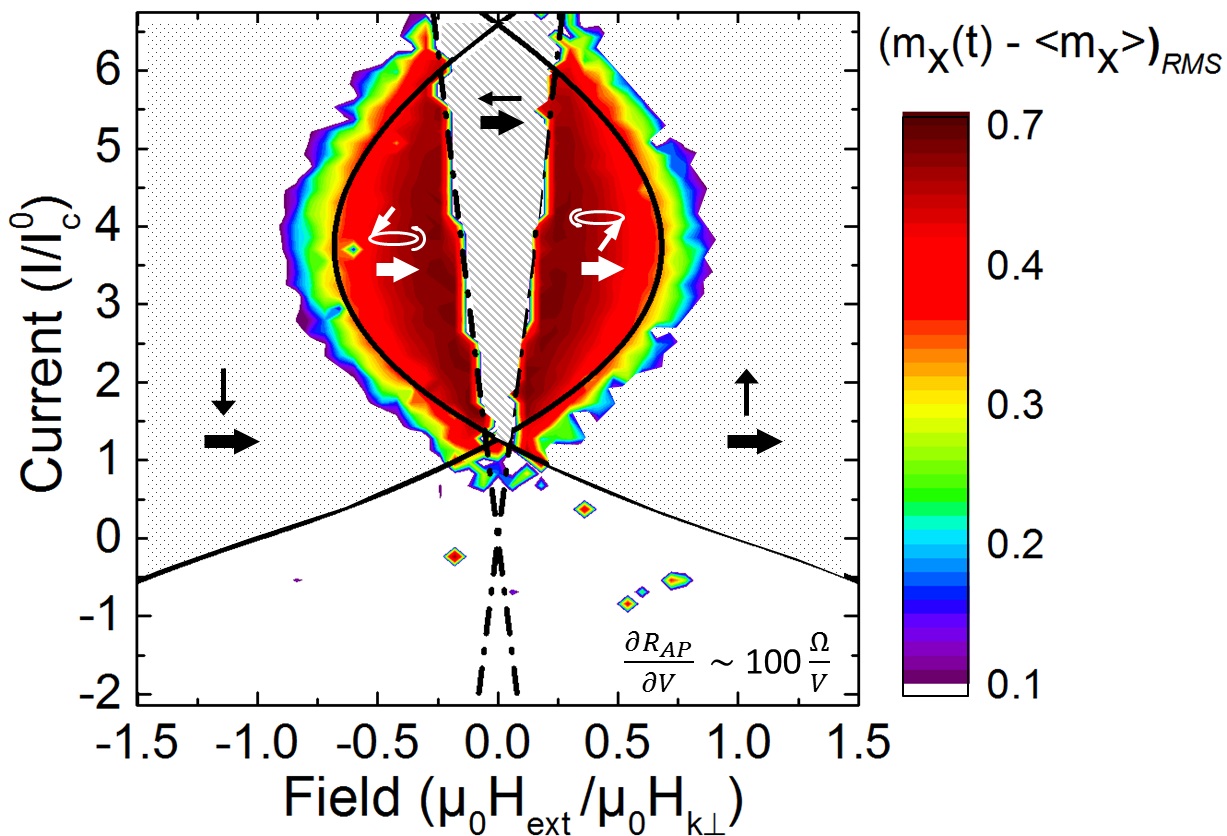}
\caption{Dynamical phase diagram of an STNO with hybrid geometry for the case when both the angular and bias dependencies of TMR are included ($\partial R_{AP}/\partial V$ is taken as 100\,$\Omega / V$). The diagram shows results of numerical integration (dynamics occurs in the coloured areas) and analytically determined onset currents (solid lines). The dash-dot lines (see eq.~\ref{eq_6_1}) defining the stability area of the~static in-plane AP state (striped area) remain unchanged compared to the case when $\partial R_{AP}/\partial V$\,=\,0\,$\Omega / V$. The solid lines (see eq.~\ref{eq_xxx}) are boundaries between the region of OOP dynamics (coloured areas) and the static OOP state (dotted areas), and exhibit a specific bended shape reflecting a gradual quenching of dynamics with an increasing current for currents above $I/I_c^0$\,=\,3.5. Current values, $I$, are normalized by $I_c^0$, i.e. the current value at the crossing of the critical lines for dynamics for the case of $\partial R_{AP}/\partial V\,=\,0$ (the crossing point of the solid lines in Fig.~\ref{Diagram_dRdV_0}). The magnetic field $\mu_0 H_{ext}$ is normalized by the effective out-of-plane anisotropy of the free layer ($\mu_0 H_{k_{\bot}}$). Magnetic configurations corresponding to static and dynamic states are marked with black and white arrows, respectively.}
\label{Diagram_dRdV_100}
\end{figure}

In order to prove that the solutions \ref{eq_xxx} and \ref{eq_6_1_bias}, indeed, determine the boundaries of steady-state precession region, the analytical results were double-checked with numerical simulation data.
Fig.~\ref{Diagram_dRdV_100} shows consistent results of the analytical calculations (solid and dash-dot lines) and numerical integration (colour scale representing the magnitude of intensity of magnetization dynamics) for the case when the bias dependence of magnetoresistance is taken into account. In the simulations, we used the same parameters as in section~\ref{Angular dependence of the TMR as a mechanism for sustained precession} and, additionally, a~realistic value of the proportionality constant defining the linear bias dependence of the~TMR, i.e. $\partial R_{AP} / \partial V$\,=\,100\,$\Omega / V$.

According to the results shown in Fig.~\ref{Diagram_dRdV_100}, for currents between $I/I_c^0$\,=\,1 and $I/I_c^0$\,=\,3.5, the diagram exhibits a non-linear (quasi-parabolic) increase of the critical current for dynamics as a function of the applied field, while for currents above $I/I_c^0$\,=\,3.5, the~boundaries of the dynamical region are bending towards each other while increasing an applied DC current and, finally, cross at $I/I_c^0$\,=\,6.5. 
Consequently, the dynamical area is now significantly reduced compared to the extended dynamical region presented in Fig.~\ref{Diagram_dRdV_0}, determined for the case when the bias dependence of the resistance is neglected, i.e. for $\partial R_{AP} / \partial V$\,=\,0\,$\Omega / V$. Moreover, above a certain current value (here, above $I/I_c^0$\,=\,3.5), any further increase of the magnitude of the DC current leads to a~decrease of the intensity of magnetization dynamics, as well as the reduction of the dynamical region (coloured area).
This effect, opposite to the one observed for the case when $\partial R_{AP} / \partial V$\,=\,0\,$\Omega / V$ (see Fig.~\ref{Diagram_dRdV_0}), is undesirable from the point of view of potential STNO applications. Thus, the minimization of the $\partial R_{AP} / \partial V$ parameter, which is mostly a material parameter of MTJ stacks, should be one of the priority aspects while designing a final commercial device, in order to limit the output power loss at large applied currents.
The pnenomenon of quenching of dynamics at large applied currents in hybrid geometry STNOs was also observed by us experimentally (see Ref.[23]).

This effect, opposite to the one observed for the case when $\partial R_{AP} / \partial V$\,=\,0\,$\Omega / V$ (see Fig.~\ref{Diagram_dRdV_0}), is undesirable from the point of view of potential STNO applications. Thus, the minimization of the $\partial R_{AP} / \partial V$ parameter, which is mostly a material parameter of MTJ stacks, should be one of the priority aspects while designing a final commercial device, in order to limit the output power loss at large applied currents.
A similar behaviour (i.e., quenching of the dynamics for large applied currents) was also observed by us experimentally; however, a detailed comparison between theoretical and experimental data will be presented elsewhere.

\subsubsection{Precession angle}

\begin{figure}[H]
\centering
\includegraphics[width=7cm]{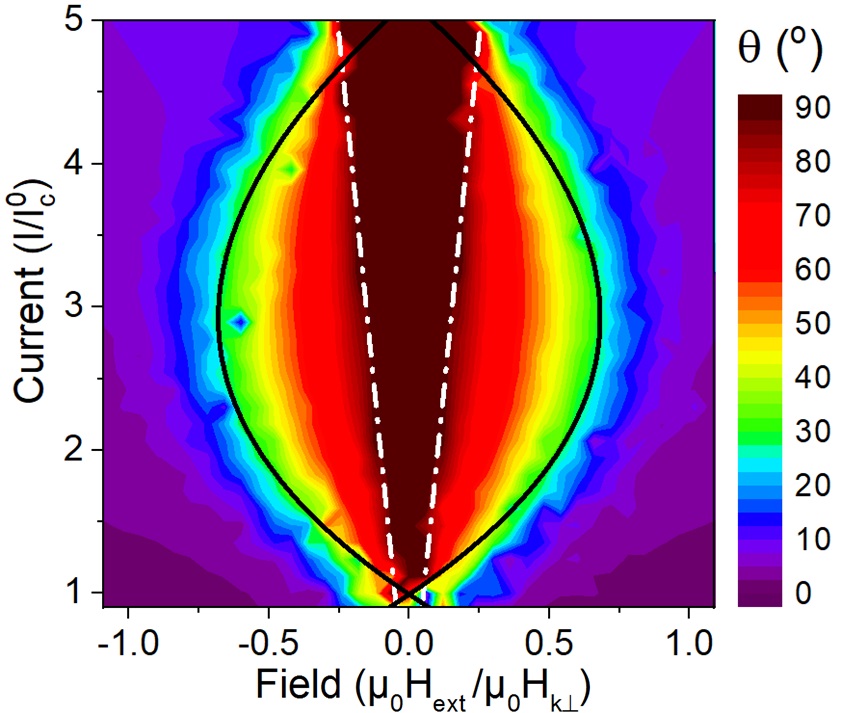}
\caption{Average magnetization precession angle $\theta$ as a function of the applied current and field for the realistic case of $\partial R_{AP}/\partial V\,\approx\,100\,\Omega/V$. Analytically determined stability region of the AP state and the critical lines for dynamics are plotted with the~white dash-dot lines (in agreement with eq.~\ref{eq_6_1}) and black solid lines (in agreement with eq.~\ref{eq_xxx}), respectively.}
\label{angle}
\end{figure}

Numerically obtained precession angle, $\theta$, as a function of current and field determined for $\partial R_{AP}/\partial V\,\approx \,100\,\Omega/V$ is shown in Fig.~\ref{angle}. Within the dynamical regions (i.e., corresponding to the colour areas in Fig.~\ref{Diagram_dRdV_100}), the precession angle increases gradually from around $10^{\circ}$ up to around $85^{\circ}$, when approaching the stability region of the~AP state (i.e., when moving closer to the white dash-dot lines).
Regarding to stable static states in the system, the stability regions of static OOP states (corresponding to the~dotted areas in Fig.~\ref{Diagram_dRdV_100}) are characterized by the precession angle close to $0^{\circ}$ (purple region outside the solid lines), while the stability region of static AP state (marked with the~striped area in Fig.~\ref{Diagram_dRdV_100}), is represented by the $90^{\circ}$ angle (dark brown area between the~dash-dot lines).

\begin{figure}[H]
\centering
\includegraphics[width=6.5cm]{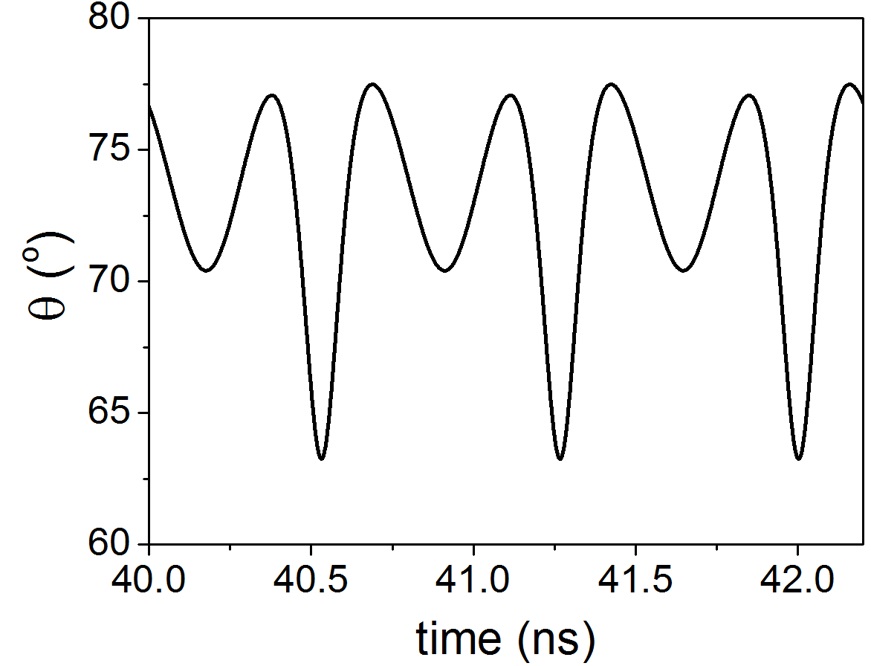}
\caption{Variation of the magnetization precession angle versus time for the case of $\partial R_{AP}/\partial V$\,=\,100\,$\Omega/V$ for $I/I_c$\,=\,3 and $\mu_0 H/\mu_0 H_{k_{\bot}}$\,=\,0. The local minima reaching 63$^{\circ}$ and 71$^{\circ}$ correspond to $\varphi$\,=\,0 (P state) and $\varphi$\,=\,$\pi$ (AP state), respectively.}
\label{theta_vs_time}
\end{figure}

Fig.~\ref{angle} clearly shows a discrepancy between the analytical solution for stable OOP dynamics (i.e., solid lines) and the numerical results (colour areas). Namely, while the solid lines define boundaries of the dynamical region where the precession angle approaches zero ($\theta\,\rightarrow\,0^{\circ}$), in the numerical data, the precession angle at these lines reaches already around $30^{\circ}$. Consequently, the analytically obtained onset fields for precession, marked with the black solid lines, are underestimated.
One of the reasons of this discrepancy may be the assumption of a constant precession angle $\theta$ for a given applied current (used in the analytical calculations), since according to the numerical results (which, in fact, reproduce the magnetization motion in a more realistic way), the precession angle deviates by around 10$^{\circ}$ when the magnetization changes its position from $\varphi$\,=\,0 to $\varphi$\,=\,$\pi$ (to be more specific, the precession cone is tilted towards $-x$ direction).
For instance, for the case of $\partial R_{AP}/\partial V$\,=\,100\,$\Omega/V$ for $I/I_c$\,=\,3 and $\mu_0 H/\mu_0 H_{k_{\bot}}$\,=\,0.2, the angle reaches the maximum of 71$^{\circ}$ close to the AP state (where the STT$_{\|}$ opposes the damping torque) and the minimum of 63$^{\circ}$ close to the P state (where the STT$_{\|}$ acts like the damping torque), as shown in Fig.~\ref{theta_vs_time}. It is, then, important to note that the magnetization precession angle plotted in Fig.~\ref{angle} is just an~average value calculated from all instant positions of the magnetization within the last 2\,ns of the simulation.

\subsubsection{The case of a large $\partial R_{AP}/\partial V$ constant}

\begin{figure*}[!ht]
\centering
\includegraphics[width=\textwidth]{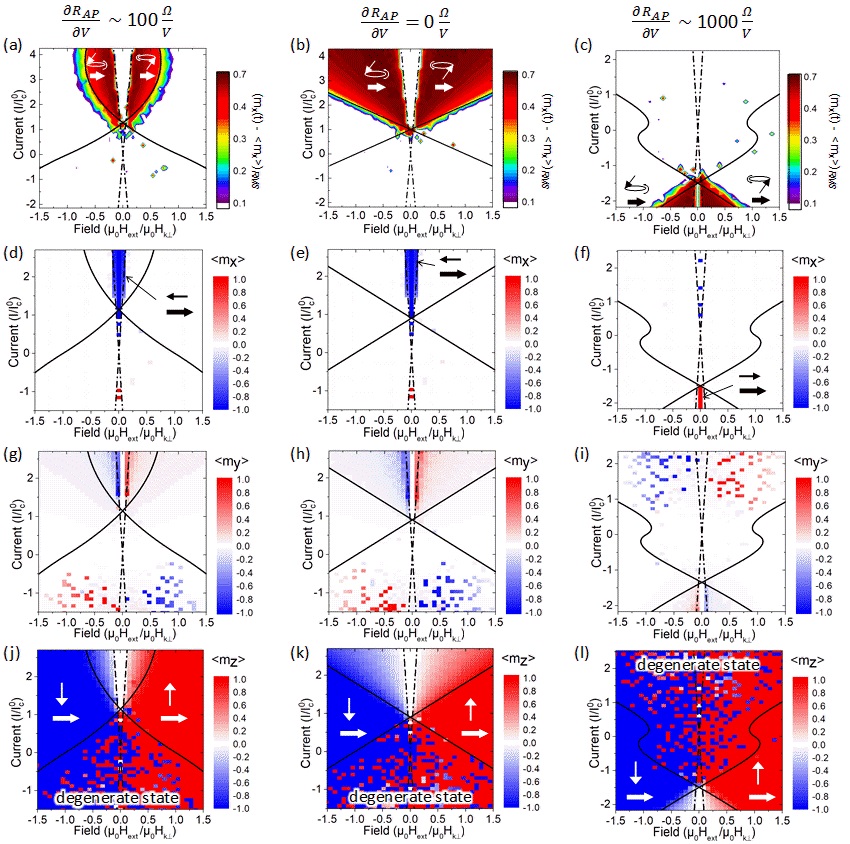}
\caption{Dynamic and static diagrams of the STNO with considered geometry for the case where both the angular and the bias dependencies of TMR are included: for the realistic case of $\partial R_{AP}/\partial V$\,$\sim$\,100\,$\Omega/V$ ((a), (d), (g) and (j)), for the limit of $\partial R_{AP}/\partial V$\,$\rightarrow$\,0\,$\Omega/V$ ((b), (e), (h) and (k)), and for the limit of a large $\partial R_{AP}/\partial V$ parameter, $\partial R_{AP}/\partial V$\,=\,1000\,$\Omega/V$ ((c), (f), (i) and (l)). Diagrams show results of numerical integration: colored areas representing intensity of magnetization dynamics in (a), (b), and (c); red and blue areas showing the average $m_x$ ((d), (e), (f)), $m_y$ ((g), (h), (i)), and $m_z$ ((j), (k), (l)) components, indicating a presence of given static states. Analytically determined onset currents for precession are plotted with black solid lines, and analytically determined borders of stability regions of the in-plane P/AP states are plotted with dash-dot lines. Current values, $I$, are normalized by $I_c^0$, i.e. the current value at the crossing of the critical lines for dynamics for the case of $\partial R_{AP}/\partial V\,=\,0$ (the crossing point of the solid lines in Fig.~\ref{Diagram_dRdV_0}).  Field values $\mu_0 H_{ext}$ are normalized by the effective out-of-plane anisotropy $\mu_0 H_{k_{\bot}}$. Corresponding magnetic configurations of static states are marked with arrows.}
\label{fig_3_g}
\end{figure*}

Within the previous paragraphs, the dynamic and static states in hybrid geometry MgO-MTJs are described for the two cases: the realistic case, where the bias dependence of the TMR is included (for $\partial R_{AP}/\partial V$\,$\sim$\,100\,$\Omega/V$), and the ideal case, where the bias dependence of the resistance is neglected (i.e., in the limit of $\partial R_{AP}/\partial V$\,$\rightarrow$\,0). Now, let us move the $\partial R_{AP}/\partial V$ parameter to the hypothetical limit of a large number, i.e. $\partial R_{AP}/\partial V$\,$\sim$\,1000\,$\Omega/V$, and use it in equation~\ref{eq_xxx} and in the equivalent numerical simulation.

All three cases are presented in Fig.~\ref{fig_3_g}. The results of the numerical and analytical calculations for $\partial R_{AP}/\partial V$\,=\,100\,$\Omega/V$ and $\partial R_{AP}/\partial V$\,=\,0\,$\Omega/V$ are recalled for the~comparison in the first and the second column, respectively. The third column contains a~corresponding dynamical diagram (Fig.~\ref{fig_3_g}(c)), as well as static in-plane (Fig.~\ref{fig_3_g}(f) and (i)) and static out-of-plane diagrams (Fig.~\ref{fig_3_g}(l)) for $\partial R_{AP}/\partial V$\,=\,1000\,$\Omega/V$. According to these diagrams, in the limit of a large $\partial R_{AP}/\partial V$ constant, the OOP dynamics occur for the opposite current sign (represented with coloured area in Fig.~\ref{fig_3_g}(c)), i.e. for electrons flowing from the reference to the free layer, favouring the P state. Due to the change of the current polarity from negative to positive, the gap in the dynamics at zero- and low fields is now the stability region of the in-plane P state (see the red area in Fig.~\ref{fig_3_g}(f)). Similarly to the cases of $\partial R_{AP}/\partial V$\,=\,100\,$\Omega/V$ and $\partial R_{AP}/\partial V$\,=\,0\,$\Omega/V$ (Fig.~\ref{fig_3_g}(j) and (k), respectively), the static OOP states (red and blue areas in Fig.~\ref{fig_3_g}(l)) are stabilized outside the dynamical region, while static states in the overlapping area of the two static OOP states are degenerate.

The static diagram showing an average $m_y$ component, shown in Fig.~\ref{fig_3_g}(i), exhibits some similarities to the graphs presented in Fig.~\ref{fig_3_g}(g) and Fig.~\ref{fig_3_g}(h); namely, a non-zero $m_y$ component in the area of "degenerate static state" (marked in Fig.~\ref{fig_3_g}(l)), as well as a small $m_y$ component in the dynamical region (marked in Fig.~\ref{fig_3_g}(c)), which indicates a slight tilt of the magnetization precession cone (see description to Fig.~\ref{Diagram_dRdV_0_static}(b)).

In order to find the reason of a presence of dynamics for the opposite current sign in the limit of a large $\partial R_{AP}/\partial V$ constant, the analytical solutions for different $\partial R_{AP}/\partial V$ parameters were analyzed. It was found that, in the~case of $\partial R_{AP}/\partial V$\,$\rightarrow$\,1000\,$\Omega/V$, the slope of the bias dependence of the $R_{AP}$ is so steep that the onset current for precession is already in the voltage range where $R_{AP}\,<\,R_{P}$, which results in a change of the TMR sign from positive to negative, and eventually leads to a stabilization of dynamics for the opposite current polarity.
To make it clear, let us go back to our initial considerations shown in the scheme in Fig.~\ref{fig_1}, which are based on the assumption that the spin-transfer torque overcomes the damping on the half of the precession trajectory where the MTJ resistance is larger (i.e., close to the AP configuration). Namely, for the most common case of the positive TMR (where $R_{AP}\,>\,R_P$), shown Fig.~\ref{fig_1}(b), STT overcomes the~damping for the ${\bf m}$ vector approaching the AP state (i.e., dynamics occur for $I_{DC} > 0$), while for the case of the negative TMR (where $R_{AP} < R_P$), STT overcomes the~damping on the other half of the precession trajectory, i.e. close to the P state (where dynamics occur for $I_{DC}$~$<$~$0$). 

Numerical results obtained for $\partial R_{AP}/\partial V$\,$\rightarrow$\,1000\,$\Omega/V$ represent the general case of negative tunnel magnetoresistance. Negative TMR ratios has been experimentally measured in magnetic tunnel junctions with specific compositions, like in TMR multilayers based on LSMO~\cite{Teresa_1999,Pantel_2012} or Mn-Ga Heusler alloys~\cite{Ma_2012,Titova_2019}. In hybrid geometry STNOs based on such materials, one should then expect dynamics to occur for opposite current sign compared to transition metal-based MTJs. It is also worth to note that STNO devices based on Mn-Ga compounds exhibit a non-trivial bias dependence of TMR, where TMR changes a sign as a function of applied voltage~\cite{Borisov_2016,Titova_2019}, which my lead to a specific dynamic characteristis of such devices, e.g. where dynamics are allowed for both current directions.

\subsubsection{Influence of STNO parameters}

\begin{figure}[H]
\centering
\includegraphics[width=8cm]{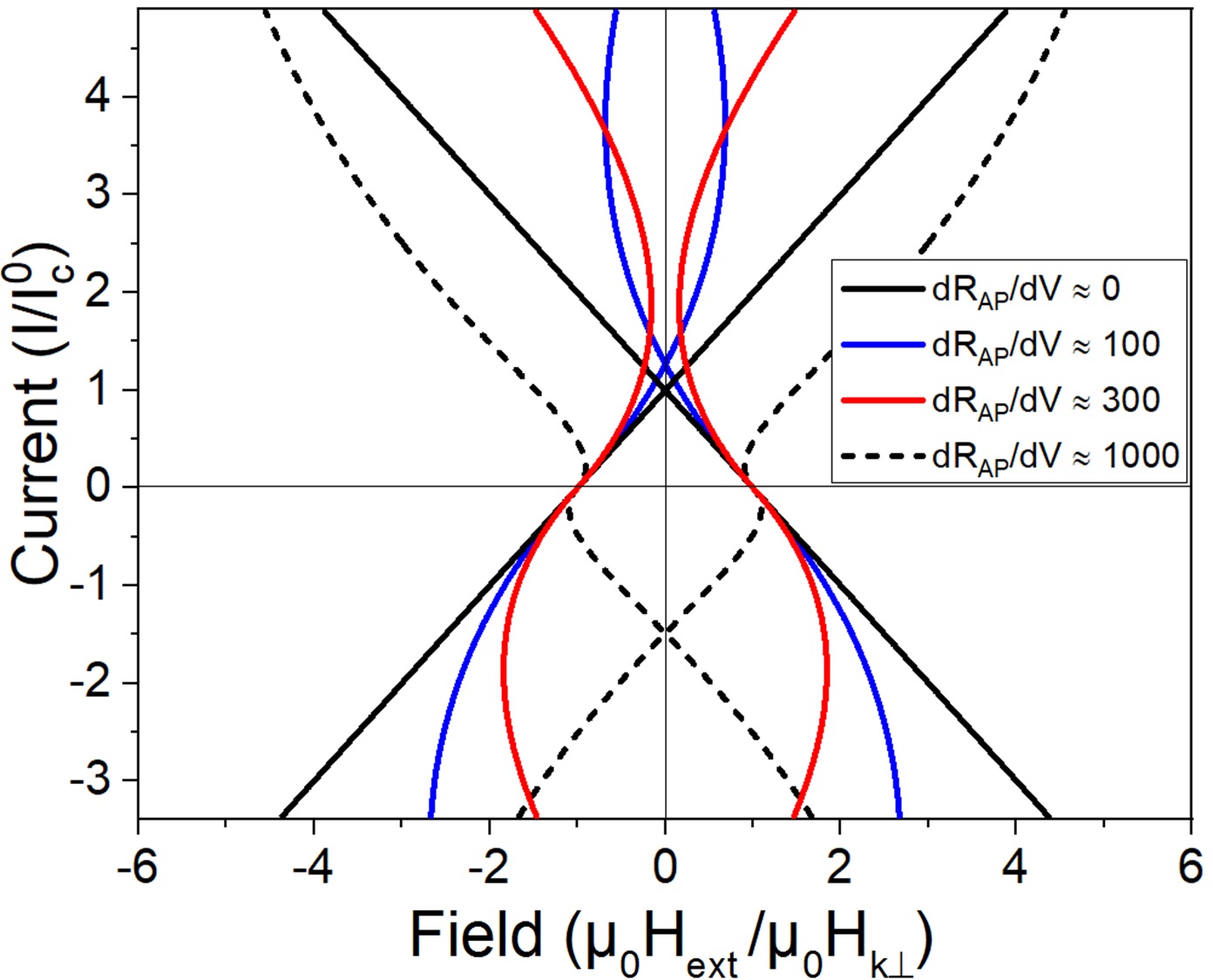}
\caption{The critical lines for STNO dynamics (see eq.~\ref{eq_xxx}) for different values of the $\partial R_{AP}/\partial V$ constant.}
\label{asymmetry_1}
\end{figure}

Fig.~\ref{asymmetry_1} shows how the magnitude of $\partial R_{AP}/\partial V$ constant influences the critical lines for STNO dynamics. For the case of $\partial R_{AP}/\partial V$\,=\,0 (black lines), the dynamical region (i.e., a triangular area in the region between the lines at the positive current range) is the largest. An increase of the $\partial R_{AP}/\partial V$ up to around 100\,$\Omega/V$ (see blue lines) brings about a significant decrease of the dynamical region. A further increase of the $\partial R_{AP}/\partial V$ constant results in a decay of dynamics (see red lines for $\partial R_{AP}/\partial V$\,=\,300\,$\Omega/V$), and eventually leads to an appearance of a dynamical region at the positive current range (see dashed lines for $\partial R_{AP}/\partial V$\,=\,1000\,$\Omega/V$).

\begin{figure*}[!ht]
\centering
\includegraphics[width=16cm]{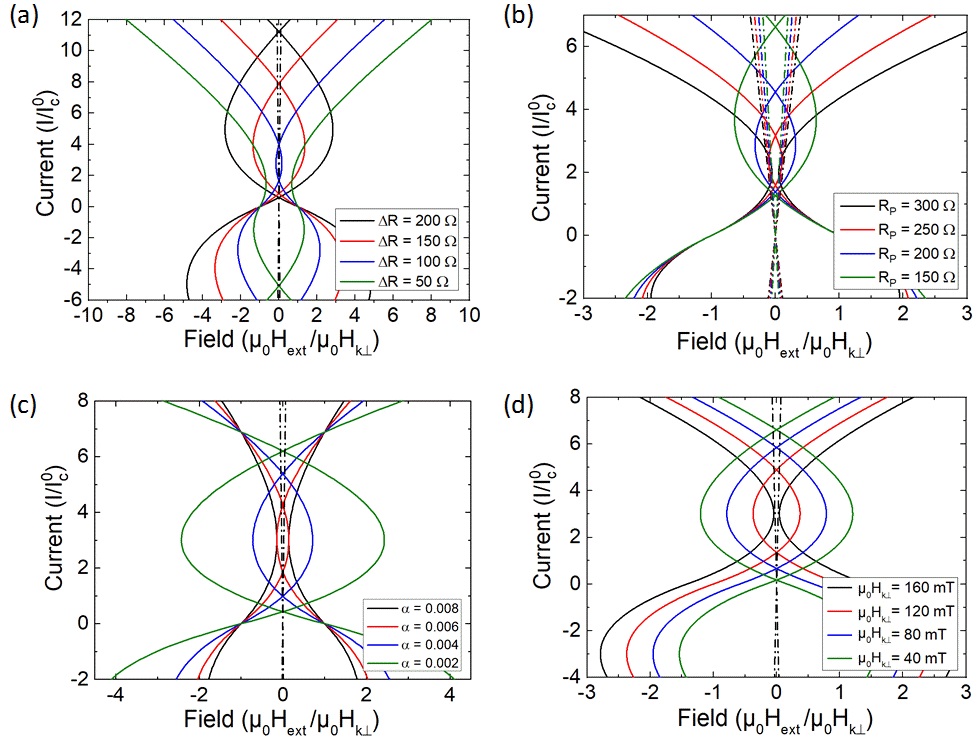}
\caption{Change in the critical lines for STNO dynamics (see eq.~\ref{eq_xxx}) as a function of following STNO parameters: (a) a resistance difference between P and AP states ($\Delta R$), (b) a resistance of the P state ($R_P$), (c) a damping constant ($\alpha$), (d) and an effective out-of-plane anisotropy ($\mu_0 H_{k_{\bot}}$). Current values, $I$, are normalized by $I_c^0$, i.e. the current value at the crossing of the critical lines for dynamics for the case of $\partial R_{AP}/\partial V\,=\,0$ (the crossing point of the solid lines in Fig.~\ref{Diagram_dRdV_0}). In the graphs (a), (b) and (c), field values $B$ are normalized by the assumed effective out-of-plane anisotropy $\mu_0 H_{k_{\bot}} = 120\,mT$. The critical lines for the stability of the static in-plane states (marked with dash-dot lines) are influenced only by the $R_P$ parameter, as shown in (b).}
\label{fig_6_g}
\end{figure*}

The influence of parameters other than $\partial R_{AP}/\partial V$ on the critical lines of the STNO diagram is shown in Fig.~\ref{fig_6_g}. Here, we consider the following parameters: a resistance difference between AP and P states ($\Delta R$), a resistance of the P state ($R_P$), a damping constant ($\alpha$), and an effective out-of-plane anisotropy ($\mu_0H_{k_{\bot}}$).
As shown in Fig.~\ref{fig_6_g}, by changing the magnitudes of these parameters, one can tune the critical currents for dynamics at given applied fields (i.e., by decreasing $\alpha$ and $\mu_0 H_{k_{\bot}}$, or by increasing $\Delta R$) or broaden the operation field range (i.e., by decreasing $\alpha$, $\mu_0 H_{k_{\bot}}$ and $R_P$, or by increasing $\Delta R$).

\subsubsection{STT angular asymmetry}

The intensity of dynamics in hybrid geometry STNOs is directly proportional to the skewness of the angular dependence of the in-plane STT component, STT$_{\|}$; namely, the larger the~deviation from a sine-type function, the more power is pumped into the system due to more effective overcoming the damping torque. Indeed, higher asymmetry of the STT$_{\|}$ angular dependence results in a larger magnetization precession angle $\theta$, which the~STNO output power is directly proportional to. According to the LLGS equation (\ref{eq_1}), the magnitude of the in-plane spin-transfer torque is a sine-type function of the angle $\beta$ between the magnetizations of the two ferromagnetic layers in the~system. However, including the angular and the bias dependence of the TMR leads to the~following expression for STT$_{\|}$:

\begin{equation}
STT_{\|}(\beta) = \frac{\partial \tau_{\|}}{\partial V} I_{DC} \frac{R_P + \frac{1}{2} \Delta R_0 (1 - \cos{\beta})}{1 + \frac{1}{2} |I_{DC}| \frac{\partial R_{AP}}{\partial V} (1 - \cos{\beta})} \sin{\beta}.
\label{eq_15_1}
\end{equation}
\\

\begin{figure}[H]
\centering
\includegraphics[width=8cm]{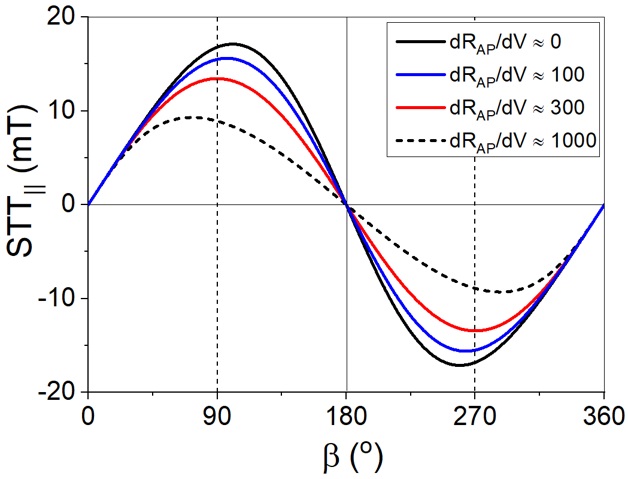}
\caption{The in-plane STT component (STT$_{\|}$) as a function of the angle between magnetizations of the free and the reference layers ($\beta$) for an applied current of $I_{DC}\,=\,1.5 \cdot I_c^0$ and different values of $\partial R_{AP}/\partial V$. Both the angular and the bias dependencies of the TMR are included.}
\label{fig_7_g}
\end{figure}

The angular dependence of the in-plane STT term for an applied current of $I_{DC}\,=\,1.5 \cdot I_c^0$ is presented in Fig.~\ref{fig_7_g}. 
The black line for $\partial R_{AP}/\partial V\,=\,0\,\Omega/V$ exhibits the spin-transfer torque asymmetry with the maximum torque at a relative angle of 102$^{\circ}$, arising solely from the cosine dependence of the resistance, when experiments are conducted at a constant applied current and the in-plane STT scales as the~corresponding voltage across the device. For this particular case, we can be observe the~highest asymmetry, which, according to Fig.~\ref{asymmetry_1} (see black critical lines), corresponds to the lowest critical currents for dynamics for $\partial R_{AP}/\partial V\,=\,0\,\Omega/V$.

Increasing the value of $\partial R_{AP}/\partial V$ to 100\,$\Omega/V$ and 300\,$\Omega/V$ (see the blue and red lines in Fig.~\ref{asymmetry_1}) shifts the maximum of STT$_{\|}$ closer to 90$^{\circ}$ (i.e., to 98$^{\circ}$ and 88$^{\circ}$, respectively). Indeed, increasing a $\partial R_{AP}/\partial V$ constant reduces the TMR amplitude at the considered applied bias and, hence, counteracts the amplitude of the cosine-type oscillations of the~resistance as a function of the angle $\beta$ between the two magnetizations, thereby decreasing the spin-torque asymmetry. For instance, for the used set of parameters, the~asymmetry disappears for $\partial R_{AP}/\partial V\,=\,330\,\Omega/V$ at $I_{DC}\,=\,1.5 \cdot I_c^0$.
It is also worth noting, that the inclusion of the bias dependence of TMR results in the reduction of the~STT$_{\|}$ angular dependence asymmetry and, simultaneously, brings about a decrease of the dynamical area in the current versus field phase diagram (see the blue lines for $\partial R_{AP}/\partial V\,=\,100\,\Omega/V$ in Fig.~\ref{asymmetry_1}).

Further increase of the $\partial R_{AP}/\partial V$ constant eventually leads to the total cancellation of the asymmetry (see a symmetric sine function for $\partial R_{AP}/\partial V$\,$\approx$\,300\,$\Omega/V$ in Fig.~\ref{fig_7_g}, and corresponding critical lines for dynamics in Fig.~\ref{asymmetry_1}).
When $\partial R_{AP}/\partial V$ is so large that the TMR becomes negative for the considered current range (i.e., $R_P$ is larger than $R_{AP}$ for currents above the onset current for precession), we can observe that the maximum of STT$_{\|}$ shifts towards the parallel configuration of the two magnetizations (i.e., towards $\beta = 0\,^{\circ}$; see the dashed line in Fig.~\ref{fig_7_g}). This results in an opening of the precession angle $\theta$ close to the parallel configuration, not to the antiparallel one (as shown in Fig.~\ref{fig_1}(b)), which results in a presence of the OOP dynamics at the negative current range (see dashed critical lines for dynamics in Fig.~\ref{asymmetry_1}).

\section{Summary}

To summarize, we present the dynamical phase diagrams of the MgO-based MTJ with IP polarizer and OOP free layer determined from the energy integral within a single magnetization precession period. We assumed that the spin-transfer torque asymmetry results from the cosine-type angular dependence of the tunnel magnetoresistance ratio, and proved that it is in fact a responsible mechanism for the precession in STNOs of this geometry. We have also determined the phase diagrams of STNO dynamics taking into account the bias dependence of the MTJ resistance. To this end, we solved the stability condition for the Jacobian matrix of out-of-plane static state, and proved that the this bias dependence exhibits drastic impact on the STNO phase diagram. With increasing slope of the AP state bias dependence ($\partial R_{AP}/\partial V$), the critical current for dynamics increases, the dynamical region is reduced and, according to the numerical integration, the intensity of the observed dynamics (i.e., output power) decreases. Indeed, the reduction of TMR due to its bias dependence suppresses the STT$_{\|}$ angular dependence asymmetry, which is in fact responsible for sustaining the precession in the spin-torque nano-oscillator. The analytical results show a very good agreement with equivalent simulation data and compare well to our previous experimental results published in Ref.[23].

\section{Acknowledgments}

E.K., V.S., C.F. and A.M.D. acknowledge support from the Helmholtz Young Investigator Initiative Grant No. VH-N6-1048.

\newpage


\begin{thebibliography}{99}


\bibitem{Slonczewski_1996}
J. C. Slonczewski, J. Magn. Magn. Mater. {\bf 159}, L1-L7 (1996).

 \bibitem{Berger_1996}
L. Berger, Phys. Rev. B {\bf 54}, 9353-9358 (1996).

 \bibitem{Tsoi_1998}
M. Tsoi, A. G. M. Jansen, J. Bass, W.-C. Chiang, M. Seck, V. Tsoi and P. Wyder, Phys. Rev. Lett. {\bf 80}, 4281 (1998).

 \bibitem{Myers_1999}
E. B. Myers, D. C. Ralph, J. A. Katine, R. N. Louie and R. A. Buhrman, Science {\bf 285}, 867-870 (1999).

 \bibitem{Katine_2000}
J. A. Katine, F. J. Albert, R. A. Buhrman, E. B. Myers and D. C. Ralph, Phys. Rev. Lett. {\bf 84}, 3149 (2000).

 \bibitem{Kiselev_2003}   
S. I. Kiselev, J. C. Sankey, I. N. Krivorotov, N. C. Emley, R. J. Schoelkopf, R. A. Buhrman and D. C. Ralph, Nature {\bf 425}, 380 (2003).

 \bibitem{Petit_2007}   
S. Petit, C. Baraduc, C. Thirion, U. Ebels, Y. Liu, M. Li, P. Wang and B. Dieny, Phys. Rev. Lett. {\bf 98}, 077203 (2007).

 \bibitem{Deac_2008}
A. M. Deac, A. Fukushima, H. Kubota, H. Maehara, Y. Suzuki, S. Yuasa, Y. Nagamine, K. Tsunekawa, D. D. Djayaprawira and N. Watanabe, Nat. Phys. {\bf 4}, 308 (2008).

 \bibitem{Villard_2010}
P. Villard, U. Ebels, D. Houssameddine, J. Katine, D. Mauri, B. Delaet, P. Vincent, M.-C. Cyrille, B. Viala, J.-P. Michel {\it et al.}, IEEE Journal of Solid-State Circuits {\bf 45}, 214-223 (2010).

 \bibitem{Choi_2014}
H. S. Choi, S. Y. Kang, S. J. Cho, I.-Y. Oh, M. Shin, H. Park, C. Jang, B.-C. Min, S.-I. Kim, S.-Y. Park {\it et al.}, Sci. Rep. {\bf 4}, 5486 (2014).

 \bibitem{Rippard_2004}
W. H. Rippard, M. R. Pufall, S. Kaka, S. E. Russek and T. J. Silva, Phys. Rev. Lett. {\bf 92}, 027201 (2004).

 \bibitem{Locatelli_2014}
N. Locatelli, V. Cros and J. Grollier, Nat. Mater. {\bf 13}, 11-20 (2014).

 \bibitem{Zeng_2013}
    Z. Zeng, G. Finocchio, B. Zhang, P. K. Amiri, J. A. Katine, I. N. Krivorotov, Y. Huai, J. Langer, B. Azzerboni, K. L. Wang and H. Jianget, Sci. Rep. {\bf 3}, 1426 (2013). 

 \bibitem{Rippard_2010}
    W. H. Rippard, A. M. Deac, M. R. Pufall, J. M. Shaw, M. W. Keller, S. E. Russek, G. E. W. Bauer and C. Serpico, Phys. Rev. B {\bf 81}, 014426 (2010).

 \bibitem{Kubota_2013}
    H. Kubota, K. Yakushiji, A. Fukushima,S. Tamaru, M. Konoto, T. Nozaki, S. Ishibashi, T. Saruya, S. Yuasa, T. Taniguchi, H. Arai and H. Imamura, Appl. Phys. Express {\bf 6}, 103003 (2013).
		
 \bibitem{Maehara_2013}
    H. Maehara, H. Kubota, Y. Suzuki, T. Seki, K. Nishimura, Y. Nagamine, K. Tsunekawa, A. Fukushima, A. M. Deac, K. Ando and S. Yuasa, Appl. Phys. Express  {\bf 6}, 113005 (2013).
		
\bibitem{Mangin_2006}
    S. Mangin, D. Ravelosona, J. Katine, M. Carey, B. Terris and E. E. Fullerton, Nat. Mater. {\bf 5}, 210–215 (2006).

 \bibitem{Taniguchi_2013}
    T. Taniguchi, H. Arai, S. Tsunegi, S. Tamaru, H. Kubota and H. Imamura, Appl. Phys. Express  {\bf 6}, 123003 (2013).

 \bibitem{Skowronski_2012}
    W. Skowronski, T. Stobiecki, J. Wrona, G. Reiss and S. Van Dijken, Appl. Phys. Express  {\bf 5}, 063005 (2012).

 \bibitem{Slonczewski_2002}
    J. C. Slonczewski, J. Magn. Magn. Mater. {\bf 247}, 324 (2002).

 \bibitem{Slonczewski_2005}
    J. C. Slonczewski, Phys. Rev. B {\bf 71}, 024411 (2005).

 \bibitem{Slonczewski_2007}
    J. C. Slonczewski and J. Z. Sun, J. Magn. Magn. Mater. {\bf 310}, 169 (2007).

 \bibitem{Kowalska_2019}
E. Kowalska, A. Fukushima, V. Sluka, C. Fowley, A. K\'{a}kay, Y. Aleksandrov, J. Lindner, J. Fassbender, S. Yuasa, and A. M. Deac, Sci. Rep. {\bf 9}, 9541 (2019).

 \bibitem{Slonczewski_1989}
    J. C. Slonczewski, Phys. Rev. B {\bf 39}, 6995 (1989).

 \bibitem{Moodera_1996}
    J. S. Moodera and L. R. Kinder, J. Appl. Phys. {\bf 79}, 8 (1996).

 \bibitem{Kubota_2008}
    H. Kubota, A. Fukushima, K. Yakushiji, T. Nagahama, S. Yuasa, K. Ando, H. Maehara, Y. Nagamine, K. Tsunekawa, D. D. Djayaprawira, N. Watanabe and Y. Suzuki, Nat. Phys. {\bf 4}, 37-41 (2008).

 \bibitem{Theodonis_2006}
    I. Theodonis, N. Kioussis, A. Kalitsov, M. Chshiev and W. H. Butler, Phys. Rev. Lett. {\bf 97}, 237205 (2006).

 \bibitem{Gao_2007}
    L. Gao, X. Jiang, S.-H. Yang, J. D. Burton, E. Y. Tsymbal and S. S. P. Parkin, Phys. Rev. Lett. {\bf 99}, 226602 (2007).

 \bibitem{Kalitsov_2013}
    A. Kalitsov, P.-J. Zermatten, F. Bonell, G. Gaudin, S. Andrieu, C. Tiusan, M. Chshiev and J. P. Velev, J. Phys.: Condens. Matter {\bf 25}, 496005 (2013).

 \bibitem{Zhang_1997}
    S. Zhang, P. M. Levy, A. C. Marley and S. S. P. Parkin, Phys. Rev. Lett. {\bf 79}, 3744 (1997).

 \bibitem{Han_2001}
    X.-F. Han, A. C. C. Yu, M. Oogane, J. Murai, T. Daibou and T. Miyazaki, Phys. Rev. B {\bf 63}, 224404 (2001).

 \bibitem{Yuasa_2004}
S. Yuasa, T. Nagahama, A. Fukushima, Y. Suzuki and K. Ando, Nat. Mater. {\bf 3}, 868-871 (2004).

 \bibitem{Coey_2010}
J. M. D. Coey, Cambridge University Press (2010).

 \bibitem{Guo_2015}
    Y.-Y. Guo, H.-B. Xue and Z.-J. Liu, AIP Advances, {\bf 5}, 057114 (2015).

 \bibitem{Fowley_2014}
F. Ciar{\'a}n, V. Sluka, K. Bernert, J. Lindner, J. Fassbender, W. H. Rippard, M. R. Pufall, S. E. Russek and A. M. Deac, Appl. Phys. Express {\bf 7}, 043001 (2014).

 \bibitem{Arai_2014}
H. Arai and H. Imamura, Appl. Phys. Express {\bf 7}, 023007 (2014).

 \bibitem{Teresa_1999}
    J. M. De Teresa, A. Barth\'el\'emy, A. Fert, J. P. Contour, F. Montaigne and P. Seneor, Science {\bf 286}, 507-509 (1999).

 \bibitem{Pantel_2012}
    D. Pantel, S. Goetze, D. Hesse and M. Alexe, Nat. Mater. {\bf 11}, 289-293 (2012).

\bibitem{Ma_2012}
    Q. L. Ma, T. Kubota, S. Mizukami, X. M. Zhang, H. Naganuma, M. Oogane, Y. Ando and T. Miyazaki, Appl. Phys. Lett. {\bf 101}, 032402 (2012).

\bibitem{Titova_2019}
     A. Titova, C. Fowley, E. Clifford, Y.-C. Lau, K. Borisov, D. Betto, G. Atcheson, R. H\"{u}bner, C. Xu, P. Stamenov, M. Coey, K. Rode, J. Lindner, J. Fassbender and A. M. Deac, Sci. Rep. {\bf 9}, 4020 (2019).

\bibitem{Borisov_2016}
    K. Borisov, D. Betto, Y.-C. Lau, C. Fowley, A. Titova, N. Thiyagarajah, G. Atcheson, J. Lindner, A. M. Deac, J. M. D. Coey, P. Stamenov and K. Rode, Appl. Phys. Lett. {\bf 108}, 192407 (2016).
    
\end{thebibliography}
\end{document}